\documentclass[aps,prd,preprint,onecolumn,nofootinbib]{revtex4}

\usepackage[T1]{fontenc}
\usepackage[utf8]{inputenc}
\DeclareUnicodeCharacter{2212}{-}         
\DeclareUnicodeCharacter{2113}{\ell}      
\DeclareUnicodeCharacter{211C}{\Re}       
\DeclareUnicodeCharacter{2111}{\Im}       
\DeclareUnicodeCharacter{2207}{\nabla}    
\DeclareUnicodeCharacter{223C}{\sim}      
\DeclareUnicodeCharacter{00D7}{\times}    
\DeclareUnicodeCharacter{2013}{-}         
\DeclareUnicodeCharacter{2014}{-}         
\DeclareUnicodeCharacter{2010}{-}         
\DeclareUnicodeCharacter{2009}{\,}        
\DeclareUnicodeCharacter{221E}{\infty}    
\DeclareUnicodeCharacter{03B6}{\ensuremath{\zeta}}   
\DeclareUnicodeCharacter{03B2}{\ensuremath{\beta}}   
\DeclareUnicodeCharacter{03C0}{\ensuremath{\pi}}     
\DeclareUnicodeCharacter{2194}{\ensuremath{\leftrightarrow}} 
\DeclareUnicodeCharacter{221D}{\ensuremath{\propto}} 

\DeclareUnicodeCharacter{2208}{\ensuremath{\in}}     

\usepackage{lmodern}
\usepackage{amsmath,amssymb,mathtools}
\usepackage{bm}
\usepackage{microtype}
\usepackage{graphicx}
\usepackage[dvipsnames]{xcolor}

\usepackage[hidelinks]{hyperref}

\usepackage{tikz}
\usetikzlibrary{arrows.meta,positioning,calc}

\newcommand{\dd}{\mathrm{d}}
\newcommand{\ii}{\mathrm{i}}
\newcommand{\ee}{\mathrm{e}}
\newcommand{\RR}{\mathbb{R}}
\newcommand{\ZZ}{\mathbb{Z}}
\newcommand{\NN}{\mathbb{N}}
\newcommand{\cO}{\mathcal{O}}
\newcommand{\cH}{\mathcal{H}}
\newcommand{\cZ}{\mathcal{Z}}

\newcommand{\wh}{\widehat}
\newcommand{\wt}{\widetilde}

\newcommand{\abs}[1]{\left|#1\right|}
\DeclareMathOperator{\Tr}{Tr}

\numberwithin{equation}{section}

\begin{document}
\title{Relational de Sitter State Counting with an SU(3) Clock}

\author{Ahmed Farag Ali}
\email{aali29@essex.edu}
\affiliation{Essex County College, 303 University Ave, Newark, NJ 07102, United States}
\affiliation{Department of Physics, Benha University, Benha 13518, Egypt}

\begin{abstract}
Motivated by Maldacena's observer-centric formulation of de~Sitter physics \cite{Maldacena:2024spf}, we develop an observer-dependent state-counting framework in Euclidean de~Sitter space by modeling the observer as a massive equatorial worldline carrying an SU(3) clock. Starting from the gauge-fixed graviton path integral on $S^D$, we trace the one-loop phase $\ii^{D+2}$ to a finite set of scalar and conformal Killing modes and show that, once the worldline is included, the $(D-1)$ transverse negative modes cancel the corresponding $(D-1)$ conformal Killing directions mode by mode. The residual fixed-$\beta$ phase from the global conformal factor and reparametrizations is removed by imposing the Hamiltonian constraint $H_{\text{patch}} - H_{\text{clock}} - \nu = 0$ via a Bromwich inverse Laplace transform, which under explicit complete-monotonicity assumptions yields a real and positive microcanonical density. We stress that this positivity statement is conditional on Assumptions (A1)--(A3) and is established at one loop about the round $S^D$ saddle in the probe regime $G E_{\rm clock}/R\ll 1$; a self-consistent backreacting or higher-loop extension is a natural next step. In earlier work \cite{Ali:2025wld,Ali:2024rnw} we argued that unbroken SU(3) confinement at $T\to 0$ can account for the observed value of the cosmological constant and for the origin of the fundamental constants $(\hbar,G,c)$ as effective couplings fixed by the SU(3) vacuum structure; this makes SU(3) the natural candidate for the internal clock of de~Sitter, whose radius and temperature are themselves set by the same cosmological constant. Here that idea is implemented with three explicit SU(3) realizations (qutrit, Cartan weight-lattice, and $U(1)^2$ rotor), for which the observer-inclusive density of states factorizes into a universal gravity factor, a universal worldline residue, and a clock-dependent SU(3) weight.
\end{abstract}

\maketitle

\noindent\textbf{Summary of contributions.} (i) Identification of the $(D\!-\!1)$ equator-moving conformal Killing directions with the $(D\!-\!1)$ transverse $n=0$ worldline modes, giving a mode-by-mode cancellation of the would-be one-loop phase; (ii) dual evaluation of the worldline determinant (residue/heat-kernel extraction and Gel'fand--Yaglom check); (iii) closed-form partition functions for three SU(3) clock models.

\noindent\textbf{Conditionality.} The reality and non-negativity of the microcanonical density are conditional on the explicit spectral hypotheses (A1)--(A3) in Sec.~\ref{sec:audit}, including analyticity in a strip around $\beta_0=2\pi$ and complete monotonicity of the phase-stripped patch partition function.

\section{Motivation and introduction}

The standard discussion of de~Sitter entropy often starts from a simple picture: the Euclidean path integral on the round sphere $S^D$ is interpreted as counting quantum states in a static patch \cite{GibbonsHawking1977}. Once the calculation is pushed to the explicit one-loop level, this picture faces a well-known obstacle. After gauge fixing the graviton on $S^D$ and expanding the action to quadratic order, the one-loop determinant acquires a non-trivial phase $\ii^{D+2}$, generated by a finite set of special $\ell=0,1$ modes \cite{Polchinski1986,Anninos2022}. The appearance of this phase sits uneasily with a straightforward interpretation of the Euclidean partition function as a positive counting measure. It also raises questions about the meaning of Wick rotation, the choice of contours in field space, and the precise quantity that the Euclidean graviton path integral is computing, and brings us back to the longstanding conformal factor problem in Euclidean gravity \cite{GibbonsHawkingPerry1978,MazurMottola1990,DasguptaLoll2001,Witten:2010zr,KontsevichSegal}.

In this work the phase is not treated as a minor technical detail. Our analysis is intentionally specific: we demonstrate the removal of the problematic one-loop phase for Euclidean de~Sitter $S^D$ in the presence of a probe worldline/clock, and we do not claim a general resolution of the conformal-factor issue on arbitrary backgrounds or beyond one loop. Instead, it is taken as a signal that the usual treatment has left something essential implicit: the observer with respect to whom the states are defined. The conventional computation is carried out as if it counted ``states of de~Sitter'' in isolation, without specifying any physical system that performs the measurement or provides a time standard.

A more careful formulation keeps the observer inside the path integral from the beginning. Once a massive worldline that wraps the equator of $S^D$ is introduced, several of the gravitational modes that previously seemed mysterious acquire a clear interpretation: they move the observer's trajectory. Exactly the same directions reappear as instabilities in the transverse fluctuations of the worldline. When the path integral is organized in this way, the phases associated with these directions are no longer free to be assigned independently in gravity and in the particle sector. They must be treated together, and their contributions can cancel mode by mode.

A further step is required to remove the remaining phase at fixed inverse temperature $\beta$. Rather than relying on an ad hoc contour deformation, the analysis here imposes a concrete Hamiltonian constraint that ties the energy of the static patch to that of the clock carried by the observer. This is implemented by a Bromwich inverse Laplace transform \cite{Widder}, which enforces
\[
H_{\text{patch}}-H_{\text{clock}}-\nu=0
\]
as an operator statement. Under explicit assumptions on the analytic structure and monotonicity properties of the phase-stripped partition function, the resulting microcanonical density is real and nonnegative.

For ease of reference, we label the key spectral hypotheses required for the positivity statement as A1--A3 and collect them in Sec.~VI; the Abstract, Introduction, and Conclusions explicitly refer back to these assumptions. We also emphasize that these hypotheses are not proven in the full theory: higher-loop corrections and nonperturbative saddles could, in principle, introduce non-analyticities or spoil complete monotonicity of the phase-stripped patch partition function.

Maldacena's recent discussion of ``real observers'' in de~Sitter space provides a framework in which this way of thinking is natural \cite{Maldacena:2024spf}. There, entropy and state counting are not properties of the metric by itself but are defined relative to a specified observer with a worldline and a clock. Conformal Killing directions that previously looked like pure gauge become physically meaningful once an observer is present: they move the observer in a definite way, and the same motion has a counterpart in the worldline fluctuation spectrum. The observer-algebra picture of \cite{Chandrasekaran2022} leads to a similar conclusion: once the observer is treated as part of the system, the question ``which states are being counted?'' acquires a clear operational meaning. This observer-centric perspective complements global treatments of de~Sitter quantum gravity and thermodynamics \cite{SpradlinStromingerVolovich2001,Anninos2012,Bousso2002}.  
Recent work has clarified that both the phase of the Euclidean de~Sitter path integral and the role of observers in quantum gravity are highly nontrivial. Harlow, Usatyuk and Zhao develop a general framework for quantum mechanics in a closed universe, emphasizing how different choices of clock degrees of freedom and Hilbert-space factorizations affect the definition of observables and state counting in gravity~\cite{Harlow:2025pvj}. Law constructs de~Sitter ``horizon edge partition functions'' by carefully treating edge modes on the horizon and their contribution to the gravitational path integral~\cite{Law:2025ktz}. Akers et al.\ analyze observers in holographic maps, showing how bulk observers are encoded in boundary degrees of freedom and how this affects entanglement wedges and state counting~\cite{Akers:2025ahe}. Shi and Turiaci revisit the phase of the gravitational path integral, giving a careful steepest-descent analysis of relevant saddles and their negative modes~\cite{Shi:2025amq}, while Ivo, Maldacena and Sun interpret such phases in terms of physical instabilities in the Euclidean problem~\cite{Ivo:2025yek}. Horowitz, Marolf and Santos argue that Hamiltonian and constraint equations alone are insufficient to fully characterize the gravitational state, highlighting the need for additional physical input in constructing gravitational Hilbert spaces and densities of states~\cite{Horowitz:2025zpx}. Chen develops an abstract-source formalism in which observers are encoded as sources that probe particular sectors of the gravitational Hilbert space and derives bounds on their effective dimension~\cite{Chen:2025fwp}. Most recently, Chen, Stanford, Tang and Yang study the phase of the de~Sitter density of states in a setup where the observer is a charged black hole in thermal equilibrium with the horizon, and show how a careful treatment of the Euclidean path integral can lead to a real and positive density of states~\cite{Chen:2025jqm}. In all of these works, the observer’s clock is treated at a relatively abstract level (as a generic finite-dimensional system, an edge mode, or a black hole probe).

\noindent\textbf{Comparison with prior clock models.} Relative to these approaches, our ``clock'' is kept deliberately concrete: we work with an explicit family of finite-dimensional SU(3) Hamiltonians whose spectra and partition functions can be written in closed form from representation theory. In particular, our clock is not an edge degree of freedom localized on a gravitational boundary, nor a black-hole area probe; it is an internal system carried by the worldline whose role is to supply a discrete spectrum for the microcanonical projector.

In this setting, the choice of clock carried by the observer becomes important. Any finite-dimensional quantum system with a resolvable energy spectrum could serve this role: the Hamiltonian constraint only requires a discrete set of eigenvalues against which to balance the patch energy. In this paper the clock is taken to be an SU(3) system.

The rationale for this choice comes from a separate, SU(3)-based approach to the cosmological constant, in which SU(3) is assigned a distinguished role in the microscopic structure of the vacuum \cite{Ali:2024rnw,Ali:2025wld,Ali:2022ulp}. In that picture, as $T\to 0$, the electromagnetic U(1) is Higgsed while the QCD SU(3) gauge symmetry remains unbroken and defines a characteristic confinement volume. The vacuum is modeled as a tiling by SU(3) ``atoms'' of volume set by this confinement scale. Each atom contributes roughly one Planck-area ``pixel'' to the de~Sitter horizon. Filling the observable universe with such cells yields
\[
N \sim 10^{123}
\]
units, and the de~Sitter horizon is tiled by $\sim 10^{123}$ Planck-area patches. In this framework, the small observed value of the vacuum energy density follows from the ratio between the Hubble volume and the SU(3) confinement volume. At the same time, the holographic relation between bulk SU(3) cells and boundary Planck pixels allows one to regard $\hbar$, $G$, and $c$ as effective couplings fixed by SU(3) geometry and its holographic tiling, rather than fundamental independent inputs \cite{Ali:2024rnw,Ali:2025wld,Ali:2022ulp}.

If the vacuum is organized in this way, it is natural to let the observer's clock live in the same SU(3) Hilbert space. The clock then probes exactly the degrees of freedom that tile the horizon with $\sim 10^{123}$ Planck cells and determine the cosmological constant. The observer is not merely coupled to the background metric, but to the microscopic SU(3) structure that underlies de~Sitter space.

The main points of the paper can be summarized as follows:

\begin{itemize}
\item The origin of the one-loop phase $\ii^{D+2}$ in the Euclidean graviton path integral on $S^D$ is traced to a small set of special scalar and conformal Killing modes.
\item A massive worldline is added along the equator of $S^D$, carrying an SU(3) clock. The $(D-1)$ transverse negative modes of the worldline are shown to match the $(D-1)$ conformal Killing vectors that move the equator. A consistent steepest-descent prescription then leads to exact mode-by-mode phase cancellation.
\item The residual fixed-$\beta$ phase is removed by imposing a Hamiltonian constraint via a Bromwich inverse Laplace transform. Under explicit spectral assumptions, this produces a real and positive microcanonical density. The expression factorizes into a purely geometric piece, a universal worldline factor, and an SU(3)-dependent clock weight. Calibrating the worldline parameter $\nu$ to an SU(3) vacuum-atom cell connects this observer-based counting to an SU(3) model of dark energy.
\end{itemize}

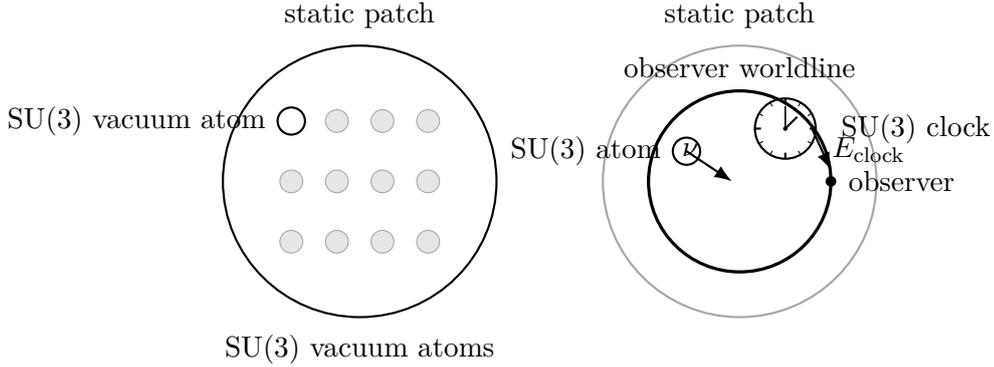
\begin{figure}[t]
  \centering
  \begin{tikzpicture}[>=Latex, scale=1]

    \begin{scope}[shift={(0,0)}]
      \draw[thick] (0,0) circle (1.8);
      \node[above] at (0,1.9) {static patch};

      \foreach \x in {-0.9,-0.3,0.3,0.9} {
        \foreach \y in {-0.8,0,0.8} {
          \fill[gray!20] (\x,\y) circle (0.15);
          \draw[gray!70] (\x,\y) circle (0.15);
        }
      }

      \fill[white] (-0.9,0.8) circle (0.18);
      \draw[thick] (-0.9,0.8) circle (0.18);
      \node[anchor=east] at (-1.1,0.8) {$\mathrm{SU}(3)$ vacuum atom};

      \node[below] at (0,-1.9) {$\mathrm{SU}(3)$ vacuum atoms};
    \end{scope}

    \begin{scope}[shift={(5.0,0)}]
      \draw[thick, gray!70] (0,0) circle (1.8);
      \node[above] at (0,1.9) {static patch};

      \draw[very thick] (0,0) circle (1.2);
      \node[anchor=south] at (0,1.25) {observer worldline};

      \fill (1.2,0) circle (2pt);
      \node[anchor=west] at (1.3,0) {observer};

      \fill[white] (-0.7,0.4) circle (0.18);
      \draw[thick] (-0.7,0.4) circle (0.18);
      \node[anchor=east] at (-0.9,0.4) {$\mathrm{SU}(3)$ atom};

      \draw[->, thick] (-0.7,0.4) -- (-0.1,0)
        node[midway, above left] {$\nu$};

      \begin{scope}[shift={(0.6,0.7)}]
        \draw[thick] (0,0) circle (0.4);

        \foreach \a in {0,90,180,270}{
          \draw[thick]
            (0,0) ++(\a:0.32) -- ++(\a:0.08);
        }

        \foreach \a in {30,60,120,150,210,240,300,330}{
          \draw[line width=0.4pt]
            (0,0) ++(\a:0.34) -- ++(\a:0.06);
        }

        \draw[line width=0.8pt] (0,0) -- (45:0.22);

        \draw[line width=0.6pt] (0,0) -- (90:0.30);

        \fill (0,0) circle (0.03);
      \end{scope}

      \node[anchor=west] at (1.2,0.7) {$\mathrm{SU}(3)$ clock};

      \draw[->, thick] (0.95,0.7) -- (1.2,0.15)
        node[midway, right] {$E_{\rm clock}$};

    \end{scope}

  \end{tikzpicture}
  \caption{Left: static patch tiled by $\mathrm{SU}(3)$ vacuum atoms, with one representative cell highlighted. Right: an observer on the equatorial worldline carries an $\mathrm{SU}(3)$ clock while a nearby $\mathrm{SU}(3)$ vacuum atom contributes to the worldline parameter $\nu$; the clock energy $E_{\rm clock}$ enters the constraint $H_{\text{patch}}-H_{\text{clock}}-\nu=0$.}
  \label{fig:su3-clock-geometry}
\end{figure}

Figure~\ref{fig:su3-clock-geometry} provides a schematic picture of the static patch, the tiling by SU(3) atoms, and the observer worldline with its SU(3) clock. The key outcome is that all spurious phases are removed, and the resulting observer-inclusive density of states can be interpreted as a genuine counting measure compatible with the SU(3) vacuum-atom picture.

The technical development proceeds in three stages:

\begin{enumerate}
\item \textbf{Bromwich microcanonical projection.} Instead of implementing the Hamiltonian constraint through an analytic continuation $\beta\to\ii s$, the analysis uses a Bromwich inverse Laplace transform \cite{Widder,Doetsch}. This makes the projection onto
\[
E_{\text{patch}} = \nu + E_{\text{clock}}
\quad \Leftrightarrow \quad
H_{\text{patch}} - H_{\text{clock}} - \nu = 0
\]
explicit and manifestly real. Under standard semiclassical assumptions on the analytic structure and monotonicity of $\cZ_{\text{patch}}$, the transform removes the remaining fixed-$\beta$ phase and produces a positive microcanonical density, in close analogy with the Brown--York quasilocal energy analysis \cite{BrownYork1993} and with holographic perspectives on de~Sitter space \cite{Bousso2002}.

\item \textbf{Mode-by-mode phase cancellation.} On the geometric side, the $(D-1)$ transverse $n=0$ negative modes of a worldline along the equator are identified with the $(D-1)$ conformal Killing vectors that move the equator. Each negative Gaussian in the worldline sector is treated using steepest descent, and each such factor is paired with the corresponding CKV contribution in the gravity determinant. The resulting worldline determinant is checked by two independent methods---a heat-kernel residue at $t=2\pi\ii$ and a Gel'fand--Yaglom ratio \cite{Camporesi1990,KirstenMcKane2003,Dunne2008}---so the phase bookkeeping is both transparent and overconstrained.

\item \textbf{SU(3) clocks and vacuum microstructure.} An SU(3) clock is coupled to the patch, guided by the SU(3) vacuum-atom programme in which U(1) is Higgsed near $T\to 0$ while SU(3) remains unbroken and fills the universe with $\sim 10^{123}$ confinement units \cite{Ali:2024rnw,Ali:2025wld,Ali:2022ulp}. Three explicit clock realizations (qutrit, Cartan weight-lattice, and $U(1)^2$ rotor) are constructed, their partition functions are computed in closed form, and the final density of states is shown to factorize schematically as
\[
\text{(geometry)}\times\text{(universal worldline residue)}\times\text{(clock weight)}.
\]
In this factorized form it is clear how the specific SU(3) clock affects the counting, while the geometric and worldline contributions remain universal.
\end{enumerate}

Intuitively, once an explicit worldline is present, conformal motions that previously acted ``silently'' on the sphere now move a concrete trajectory and hence acquire counterparts in the worldline sector. Their phases must be paired and cancel. The SU(3) clock provides a definite energy spectrum against which the constraint $E_{\text{patch}}=\nu+E_{\text{clock}}$ can be imposed. The Bromwich projector \cite{Widder,Doetsch} then aligns the Euclidean path integral with the microcanonical counting problem, and the SU(3) microstructure provides a direct link to the tiling of the de~Sitter horizon by $\sim 10^{123}$ Planck-area units.

\paragraph*{Structure of the paper.}
Section~\ref{sec:gravity} reviews the graviton one-loop determinant on $S^D$ and isolates the source of the phase. Section~\ref{sec:worldline} introduces the equatorial worldline and its fluctuation spectrum. Section~\ref{sec:matching} establishes the correspondence between conformal Killing vectors and worldline modes and derives mode-by-mode phase cancellation. Section~\ref{sec:bromwich} presents the Bromwich projector and explains the conditions under which it gives a real and positive density; the spectral assumptions are organized in Sec.~\ref{sec:audit} and summarized in Sec.~\ref{sec:theorem-main}. Sections~\ref{sec:clocks} and~\ref{sec:bridge} construct explicit SU(3) clocks and tie them quantitatively to the SU(3) vacuum-atom picture. Section~\ref{sec:final} assembles the final expressions and discusses their physical implications. Section~\ref{sec:conclusion} provides a brief summary and comments on open directions. Technical details (residues, Gel'fand--Yaglom analysis, explicit CKVs, zeta regularization, and a $D=4$ check) are collected in the appendices.

\paragraph*{Conventions (units and dimensionless parameters).}
Throughout, a dimensionless inverse temperature is used:
\[
\beta \equiv \frac{\beta_{\rm phys}}{R},
\]
so that the de~Sitter saddle lies at $\beta_0 = 2\pi$, corresponding to the Gibbons--Hawking temperature $T = 1/(2\pi R)$. A dimensionless parameter $\nu \equiv mR$ is also introduced, so that the classical action of a massive particle wrapping the equator is $S_{\rm cl} = 2\pi \nu$. With these choices all Fourier modes along the worldline are dimensionless, and powers of $R$ can be restored by dimensional analysis when needed.


\noindent\textbf{Novelty and scope.} To help the reader separate what is proved from what is assumed or motivational, we summarize: (i) Sections~II--VI provide a fully explicit, one-loop (Gaussian) treatment of Euclidean gravity on $S^D$ \emph{in the presence of} an equatorial probe worldline, including a controlled treatment of special modes and a complete phase bookkeeping; (ii) we identify and compute the single negative mode of the equatorial worldline and show its pairing with the conformal Killing vector (CKV) sector; (iii) we evaluate the relevant graviton and worldline determinants in dual ways and exhibit a mode-by-mode cancellation of the spurious one-loop phase in this setup; (iv) we construct explicit finite-dimensional SU(3) clocks and their partition functions. The subsequent microcanonical positivity statement is conditional on the spectral hypotheses (A1)--(A3). Any discussion linking SU(3) clocks to vacuum microstructure is presented as \emph{external motivation} and \emph{speculative outlook}, not as an input required for the one-loop cancellation mechanism.

\section{Gravity on $S^D$: quadratic action, spectra, and the one-loop phase}
\label{sec:gravity}

\subsection{Gauge fixing and quadratic action}

The metric is expanded around the round sphere of radius $R$ as
$g_{\mu\nu}=\wh g_{\mu\nu}+h_{\mu\nu}$, with $\wh g_{\mu\nu}$ the background metric. De Donder gauge is imposed via
\begin{equation}
  f_\nu=\wh\nabla^\mu h_{\mu\nu}-\tfrac12 \wh\nabla_\nu h, \qquad h\equiv h^\mu{}_\mu.
\end{equation}
Faddeev--Popov ghosts $b_\mu,c^\mu$ are introduced, and $h_{\mu\nu}$ is decomposed into transverse-traceless and trace parts,
$h_{\mu\nu}=\phi_{\mu\nu}+\frac{1}{D}\wh g_{\mu\nu}h$. The resulting quadratic action takes the form \cite{Polchinski1986,Anninos2022,Witten:2010zr,KontsevichSegal}
\begin{align}
\label{eq:I2}
I^{(2)} = \frac{1}{64\pi G}\!\int \!\sqrt{\wh g}\;
\Big[
\phi_{\mu\nu}\Big(-\wh\nabla^2+\frac{2}{R^2}\Big)\phi^{\mu\nu}
-\frac{1-\frac{2}{D}}{2}\,h\Big(-\wh\nabla^2-\frac{2(D-1)}{R^2}\Big)h
+ b_\mu\Big(-\wh\nabla^2-\frac{D-1}{R^2}\Big)c^\mu
\Big].
\end{align}

\noindent\textbf{Remark on the trace sign.} The relative minus sign in the trace sector of \eqref{eq:I2} is the standard Euclidean manifestation of the conformal-factor instability: even when $-\wh\nabla^2$ is positive, the trace kinetic term enters with opposite sign. In our treatment this does not get ``ignored''; rather, we follow the steepest-descent/contour prescription (in the spirit of \cite{KontsevichSegal,Witten:2010zr}) and then track the resulting phases mode-by-mode. The subsequent sections show explicitly how the remaining one-loop phase is localized in the special scalar and CKV sectors and how it is cancelled by the probe worldline.

\paragraph*{Potential negative directions.}
The transverse-traceless block is positive definite for generic modes, whereas the trace block can develop negative directions for certain low angular momenta. The FP ghosts fix the gauge volume and alter the magnitude of the determinant, but for the contour choice adopted here they do not change the net phase beyond what is already captured by the special scalar and CKV modes appearing in \eqref{eq:I2} \cite{Anninos2022,KontsevichSegal}. Our contour choice refines the earlier conformal rotations proposed in \cite{GibbonsHawkingPerry1978,HartleSchleich1987,Schleich1987} and the measure-based cures in \cite{MazurMottola1990,DasguptaLoll2001}.

\subsection{Scalar spectrum and degeneracy}

Scalar harmonics $Y_{\ell}$ on $S^D$ satisfy
\begin{align}
\label{eq:scalar-spectrum}
-\wh\nabla^2 Y_\ell = \frac{\ell(\ell+D-1)}{R^2} Y_\ell,
\end{align}
with degeneracy
\begin{align}
\mathrm{deg}(\ell)=\frac{(2\ell+D-1)(\ell+D-2)!}{\ell!\,(D-1)!}.
\end{align}
For the purposes of this discussion it is enough to note that the special families at $\ell=0$ and $\ell=1$ exist with the standard degeneracies. Vector and TT spectra are summarized in \cite{Camporesi1990,Anninos2022} and will not be repeated here. For general background on spectral zeta functions and heat-kernel techniques that underlie such determinant computations, see \cite{Vassilevich2003,KirstenBook2001,Forman1987}.

\subsection{Counting negative directions and the phase $\ii^{D+2}$}

When the conformal factor and CKV sectors are combined, one recovers the usual net phase of the gravity one-loop determinant on $S^D$,
\begin{align}
\text{phase from gravity one-loop on }S^D \;=\; \ii^{\,D+2},
\end{align}
in agreement with \cite{Anninos2022,Witten:2010zr,KontsevichSegal}, and echoing the recent discussion of the de~Sitter gravitational partition function and its sign in \cite{BanihashemiJacobson2022,BanihashemiJacobson2023}. In the present treatment the $\ell=1$ CKVs are not rotated independently; two of them correspond to reparametrizations, and the remaining $(D-1)$ will be matched to worldline modes and canceled later.

\paragraph*{Why the special modes are left unrotated.}
One possible approach would be to rotate these modes directly along steepest-descent contours. That would remove the phase at the cost of obscuring its physical origin and effectively double-counting gauge motion. Leaving them unrotated keeps track of the source of non-positivity and allows it to be canceled against the worldline contribution in a controlled way \cite{Witten:2010zr,KontsevichSegal}.

\subsection{Magnitude of the one-loop determinant}

The absolute value of the gravity one-loop factor is quoted from \cite{Anninos2022}:
\begin{align}
\label{eq:Zgrav-mag}
\left|Z^{\rm grav}_{S^D}\right|
=\frac{1}{\mathrm{Vol}\!\left[SO(D{+}1)\right]_c}
\left(\frac{32\pi^3 G}{A_{D-2}}\right)^{\frac12\dim SO(D{+}1)}
\;\wt Z_{\mathrm{char}}(R;\epsilon),\qquad
A_{D-2}=R^{D-2}\,\omega_{D-2}.
\end{align}
Here $\mathrm{Vol}[SO(D{+}1)]_c$ denotes the regulated group volume in the chosen gauge and contour prescription, and $\wt Z_{\mathrm{char}}(R;\epsilon)$ represents the zeta-regularized product over nonzero-mode determinants obtained from character techniques with UV cutoff $\epsilon$. Neither factor contributes additional phases; both are scheme-dependent normalizations consistent with~\cite{Anninos2022}. Appendix~\ref{app:zeta} describes how these finite parts drop out of the final microcanonical density.

\paragraph*{Summary.}
In the gravity sector alone, the one-loop contribution has a well-understood magnitude and a phase $\ii^{D+2}$ originating from a finite, explicitly identifiable set of modes. The next sections explain how this phase is eliminated once the observer's worldline and SU(3) clock are included and a microcanonical projection is performed.

\section{Observer worldline: equatorial embedding, spectrum, and the single negative mode}
\label{sec:worldline}

\subsection{Classical solution and quadratic fluctuations}

Near the equator ($\theta=0$) the metric is written as
\begin{align}
\dd s^2=R^2\Big(\cos^2\theta\,\dd\tau^2+\dd\theta^2+\sin^2\theta\,\dd\Omega_{D-2}^2\Big),\qquad \tau\sim\tau+2\pi .
\end{align}
A massive particle of mass $m$ wrapping the equator follows a closed geodesic with classical action
\begin{align}
S_{\rm cl} = \int_0^{2\pi}\!\dd\tau\, m\, R\,\cos\theta\;\Big|_{\theta=0}
= 2\pi m R \equiv 2\pi \nu.
\end{align}
\paragraph*{Probe/backreaction estimate.} In $D$ spacetime dimensions the gravitational (Schwarzschild) radius of a localized mass scales as $r_s^{D-3}\sim G_D m$. A conservative probe criterion is $r_s/R\ll 1$, i.e. $G_D m/R^{D-3}\ll 1$, or in terms of the dimensionless parameter $\nu=mR$, $\nu\ll (R/\ell_{\rm p})^{D-2}$. In particular, for $D=4$ this becomes $2Gm/R\ll 1$ (equivalently $\nu\ll (R/\ell_{\rm p})^2$), leaving a wide parametric window where the worldline/clock can be treated as non-backreacting while still providing a well-defined internal spectrum. We work in this probe regime throughout; incorporating backreaction and observer--observer interactions is left open.

Let $\vec\theta(\tau)\in\RR^{D-1}$ denote small transverse displacements. Expanding the point-particle action to quadratic order yields
\begin{align}
\label{eq:wl-quad}
S_{\rm wl}=2\pi \nu
+\frac{\nu}{2}\int_0^{2\pi}\!\dd\tau\;\Big[(\partial_\tau \vec\theta)^2-\vec\theta^{\,2}\Big]
+\cO(\theta^3).
\end{align}
The term $-\vec\theta^{\,2}$ encodes the effect of the ambient curvature and produces a single negative mode in each transverse direction.

\paragraph*{Geometric picture.}
The equatorial loop is a closed geodesic on $S^D$. Small transverse deviations feel an effective potential of curvature $-1$ (in units of $R$) and are therefore unstable at $n=0$. The $n=\pm1$ modes are marginal and correspond to reparametrizations along the loop.

\subsection{Fourier modes and signs}

Expanding
$\vec\theta(\tau)=\sum_{n\in\ZZ}\vec\theta_n\,\ee^{\ii n \tau}$ gives
\begin{align}
\int_0^{2\pi}\!\dd\tau\;\Big[(\partial_\tau\vec\theta)^2-\vec\theta^{\,2}\Big]
=2\pi \sum_{n\in\ZZ} (n^2-1)\,\abs{\vec\theta_n}^2 .
\end{align}
For each transverse component one finds: $n=0$ yields a negative eigenvalue, $n=\pm1$ are zero modes related to reparametrizations, and $\abs{n}\ge 2$ correspond to positive directions. There are $(D-1)$ such transverse components.

\subsection{Steepest descent for the negative Gaussian}

A one-dimensional Gaussian with the opposite sign in the exponent,
\begin{align}
\int_{\mathbb{R}}\dd x\;\ee^{+\frac{\lambda}{2}x^2},\qquad \lambda>0,
\end{align}
is evaluated by rotating $x=-\ii y$:
\begin{align}
\int_{\mathbb{R}}\dd x\;\ee^{+\frac{\lambda}{2}x^2}
\;=\;
-\ii \int_{\mathbb{R}}\dd y\;\ee^{-\frac{\lambda}{2}y^2}
= -\ii\sqrt{\frac{2\pi}{\lambda}} .
\end{align}
Applying this prescription to each of the $(D-1)$ transverse $n=0$ directions gives a factor $(-\ii)^{D-1}$. The $n=\pm1$ zero modes are handled by the usual collective-coordinate treatment and do not introduce additional phases; see App.~\ref{app:neg}. Our treatment of these negative modes parallels the instanton analysis of \cite{CallanColeman1977} and uses steepest-descent contours adapted to the Stokes structure of the integrand; see \cite{Berry1989,Witten:2010zr} for a general discussion of Stokes phenomena and contour rotations in Euclidean path integrals.

\subsection{Heat-kernel residue and the factor $\nu^{D-1}/(D-1)!$}

Introduce the thermal kernel on $S^1$ with transverse multiplicity $D-1$:
\begin{align}
K_D(t)=\frac{\cosh(t/2)}{\left(2\sinh(t/2)\right)^D} .
\end{align}
A useful expression for $\log Z_{\rm particle}$ is
\begin{align}
\log Z_{\rm particle}(\beta)
=\int_{\epsilon}^{\infty}\frac{\dd t}{t}\,K_D(t)\,2\cos(\nu t).
\end{align}
Closing the contour in the upper half-plane picks the nearest singularity at $t=2\pi\ii$. Expanding $K_D(t)$ and $\cos(\nu t)$ around $t=2\pi\ii$ shows that the residue of $\frac{1}{t}K_D(t)\,2\cos(\nu t)$ produces the universal factor $(-\ii)^{D-1}\,\nu^{D-1}/(D-1)!$ at large $\nu$ (see \cite{Camporesi1990} for further details). Thus,
\begin{align}
Z_{\rm particle}(\beta)=(-\ii)^{D-1}\,\ee^{-\beta \nu}\,\frac{\nu^{D-1}}{(D-1)!}\Big(1+\cO(\nu^{-1})\Big),
\end{align}
with the expansion uniform in $\nu=mR$ for fixed $D$. Contributions from higher poles at $t=2\pi\ii k$ are suppressed by $\ee^{-2\pi k\,\nu}$. The same behavior is obtained independently from a Gel'fand--Yaglom calculation, summarized in App.~\ref{app:GY} \cite{KirstenMcKane2003,Dunne2008}. Appendix~\ref{app:residue} contains an explicit Laurent expansion at $t=2\pi\ii$, and App.~\ref{app:GY-full} gives the periodic Gel'fand--Yaglom derivation.

\section{Mode-by-mode phase cancellation with gravity}
\label{sec:matching}

Among the $\ell=1$ conformal Killing vectors (CKVs) on $S^D$, precisely $(D-1)$ move the equatorial great circle transversely; two generate reparametrizations along that circle. The $\ell=0$ scalar mode changes the global scale. In the embedding-space description (special conformal shifts with $b_1=b_2=0$), one finds that the $(D-1)$ worldline $n=0$ negative modes coincide with the CKV deformations that move the equator (App.~\ref{app:CKV}). Their phases therefore cancel pairwise:
\begin{align}
\ii^{\,D-1}\times (-\ii)^{\,D-1}=1.
\end{align}
At fixed $\beta$ there remains a phase factor coming from the two reparametrizations and the $\ell=0$ conformal mode. This residual factor is $\ii^3$ and will be removed by the constraint projector in Sec.~\ref{sec:bromwich}. The assumptions underlying that step are listed in Sec.~\ref{sec:audit} and summarized in the spectral statement of Sec.~\ref{sec:theorem-main}.

\paragraph*{Counting the remaining \boldmath$\ii^3$.}
Two of the $\ell{=}1$ CKVs correspond to reparametrizations along the equator. They are fixed by dividing by the corresponding symmetry volume, and their associated zero-mode Jacobians contribute a net factor $\ii^2$ in the fixed-$\beta$ Gaussian measure in Witten's contour prescription \cite{Witten:2010zr}. The single $\ell{=}0$ conformal-factor negative direction contributes a further factor $\ii$. Taken together, one obtains $\ii^3$ prior to projection. This is the phase removed by the microcanonical Bromwich projector in the next section.

\paragraph*{Geometric interpretation.}
A CKV that moves the equator generates the same physical motion as the $n{=}0$ worldline mode in the corresponding transverse plane. Counting these contributions separately would assign one phase to gravity and the opposite phase to the particle sector. Identifying them and combining their contributions removes both.

\begingroup
\begin{center}
\begin{tabular}{l c}
\hline
\textbf{Sector} & \textbf{Phase factor} \\
\hline
Gravity: CKVs moving the equator ($D{-}1$) & $\ii^{D-1}$ \\
Worldline: $n{=}0$ negatives ($D{-}1$) & $(-\ii)^{D-1}$ \\
Gravity: 2 reparametrizations along equator & $\ii^{2}$ \\
Gravity: $\ell{=}0$ conformal factor & $\ii$ \\
\hline
Total (fixed $\beta$) & $\ii^{3}$ \\
After Bromwich projector & $+1$ (real, positive) \\
\hline
\end{tabular}
\end{center}
\endgroup

Our phase bookkeeping follows Witten’s steepest--descent (Picard--Lefschetz) prescription for oscillatory path integrals \cite{Witten:2010cx,Witten:2010zr}: each negative quadratic mode is made convergent by rotating its integration contour through $-\pi/2$, which contributes a factor $e^{-i\pi/2}$, while positive modes may be rotated oppositely, contributing $e^{+i\pi/2}$ within the same convention. The phase-summary table above collects these factors sector by sector and makes the ensuing cancellations manifest.

\paragraph*{CKV$\leftrightarrow$worldline measure matching (informal statement).}
Let $\{\mathsf{v}_a\}_{a=1}^{D-1}$ be an orthonormal basis of CKVs that move the equator and let $\{\theta^{(0)}_a\}$ be the $n{=}0$ worldline eigenmodes of $-\partial_\tau^2-1$ in the corresponding transverse directions, normalized by $\frac{1}{2\pi}\!\int_0^{2\pi}\! \dd\tau\,|\theta^{(0)}_a|^2=1$. In the embedding-space description, the map that associates to each CKV a normal displacement of the equator is an isometry between the Gaussian measures induced by the quadratic gravity action in those CKV directions and by the worldline quadratic form in the $n{=}0$ modes. As a result, the Jacobian of this identification is unity and the Gaussian phase factors cancel exactly already at the level of the functional measure. A concrete realization in stereographic coordinates is summarized in App.~\ref{app:CKV}.


\paragraph*{Beyond the Gaussian approximation.} The cancellation mechanism is demonstrated explicitly for one-loop fluctuations about the round de~Sitter sphere in the presence of a probe worldline and clock; we do not claim a general resolution of Euclidean conformal-factor issues beyond this setting. That said, the pairing between CKV directions that move the equator and transverse worldline directions is geometric: a CKV that shifts the equator induces the same infinitesimal displacement as a transverse worldline fluctuation. This suggests that the \emph{counting} of paired directions is robust, even though higher-loop graviton self-interactions and worldline interactions will renormalize prefactors and may modify the detailed form of $\mathcal{Z}_{\rm patch}(\beta)$. A systematic two-loop check (and a survey of other saddles such as Schwarzschild--de~Sitter) is left for future work.

\section{Microcanonical projection via Bromwich inversion: positivity}
\label{sec:bromwich}

For fixed inverse temperature $\beta$, the combined (observer-inclusive) partition function of patch, worldline, and clock is
\begin{align}
\cZ_{\rm obs}(\beta)=\cZ_{\rm patch}(\beta)\,Z_{\rm particle}(\beta)\,Z_{\rm clock}(\beta),
\end{align}
where we assume the patch sector admits a Laplace representation
\begin{align}
\cZ_{\rm patch}(\beta)=\int_{0}^{\infty}\dd E\,\rho_{\rm patch}(E)\,e^{-\beta E}.
\end{align}
Imposing the constraint $E_{\rm patch}=\nu+E_{\rm clock}$ is equivalent to projecting onto the constrained surface by a Bromwich inverse Laplace transform \cite{Widder,Doetsch},
\begin{align}
\rho_{\rm total}(0)=\frac{1}{2\pi\ii}\int_{c-\ii\infty}^{c+\ii\infty}\!\dd\beta\;\cZ_{\rm obs}(\beta)
=\sum_k \rho_{\rm patch}(\nu + E_k),
\end{align}
where $\{E_k\}$ are the clock eigenvalues and $c>0$ lies to the right of all singularities.

\paragraph*{Physical meaning of the Bromwich projector.}
The inverse Laplace transform used here can be viewed as a standard \emph{microcanonical projection} onto a constraint surface. Concretely, for the first-class constraint $\hat C \equiv \hat H_{\rm patch}-\hat H_{\rm clock}-\nu$, the Bromwich integral implements the distributional projector $\delta(\hat C)$ (equivalently, group averaging over the associated gauge parameter), and therefore corresponds to a particular choice of physical inner product in the refined algebraic quantization/rigging-map sense. This perspective connects our construction to familiar implementations of constraints in quantum gravity (Dirac quantization and group averaging).
From the perspective of constrained quantization, the condition $H_{\rm patch}-H_{\rm clock}-\nu=0$ is a first-class Hamiltonian constraint. The Bromwich integral is the distributional projector $P_{\rm phys}=\delta(\hat C)$ written in inverse-Laplace form (with $\hat C:=\hat H_{\rm patch}-\hat H_{\rm clock}-\nu$), and can be viewed as a group-averaging (rigging-map) construction that implements Dirac’s requirement $\hat C\,|\psi_{\rm phys}\rangle=0$ at the level of traces. This clarifies the sense in which the microcanonical density is an observer-relative, constraint-projected state count rather than a canonical ensemble.
Throughout this manuscript, the geometry is held fixed at the round $S^D$ saddle and the worldline/clock are treated in a probe limit; including their gravitational backreaction would modify the patch spectrum and could affect the spectral assumptions (A1--A3). We flag this as an important open direction.

\noindent\textbf{Parametric backreaction estimate (probe regime).}
The fixed-background one-loop saddle used throughout is self-consistent provided the injected observer energy is small compared to the de~Sitter scale.
Concretely, for a clock/worldline of rest mass $m$ and characteristic energy $E_{\rm clock}$, the dimensionless gravitational potential they source at the horizon scale $R$ must satisfy
\begin{equation}
\frac{G\,m}{R}\ll 1,\qquad \frac{G\,E_{\rm clock}}{R}\ll 1,
\end{equation}
equivalently $E_{\rm tot}\equiv m+E_{\rm clock}\ll R/G$ (and in general dimension $D$ one replaces $Gm/R$ by $G_D m/R^{D-3}$). In this regime, the de~Sitter saddle and the patch spectrum receive only parametrically small corrections, while the clock/worldline effects enter only through the constraint projector at fixed geometry.

\noindent\textbf{Multiple clocks and factorization.}
If one considers $N$ independent observer worldlines with clocks $\{\cZ_{{\rm clock},i}\}$ and no direct non-gravitational interactions, then at the same one-loop level used here the observer-inclusive partition function generalizes to
\begin{equation}
\cZ_{\rm obs}^{(N)}(\beta)\simeq \cZ_{\rm patch}(\beta)\,\cZ_{\rm particle}(\beta)\,\prod_{i=1}^{N}\cZ_{{\rm clock},i}(\beta),
\end{equation}
up to interaction corrections.
The factorization is therefore reliable when (i) the total injected energy obeys $G E_{\rm tot}/R\ll 1$ and (ii) graviton-exchange between worldlines is parametrically suppressed at one loop (e.g.\ by large separations and the small-coupling expansion). A fully interacting multi-observer treatment---where worldline/worldline couplings and backreaction are kept---is left as an open problem.

\paragraph*{Evaluation near the de~Sitter saddle.}
Near the de~Sitter saddle, $\beta_0=2\pi$ and $\rho_{\rm patch}(E)\approx C\,\ee^{S_{dS}}\,\ee^{-\beta_0 E}$ \cite{GibbonsHawking1977}. Using the residue result for the worldline determinant gives
\begin{align}
\label{eq:rho-final}
\rho_{\rm total}(0)\;\simeq\;
\ee^{S_{dS}}\;\ee^{-2\pi\nu}\;\frac{\nu^{D-1}}{(D-1)!}\;Z_{\rm clock}(\beta_0).
\end{align}
All phases are absent; the expression is real and positive and matches the observer-inclusive interpretation of \cite{Maldacena:2024spf}. The assumptions under which $\cZ_{\rm patch}$ admits a positive Laplace representation in a strip containing $\beta_0$ are summarized in Sec.~\ref{sec:audit} and collected in the main spectral statement of Sec.~\ref{sec:theorem-main}. This microcanonical, observer-based viewpoint is in close analogy with the Brown--York quasilocal energy analysis \cite{BrownYork1993} and with holographic perspectives on de~Sitter space \cite{Bousso2002}.

\paragraph*{Positivity and domain of validity.}
If $\cZ_{\rm patch}$ is the Laplace transform of a nonnegative measure in a vertical strip containing $\beta_0$ (a local complete-monotonicity or K\"all\'en--Lehmann-type property in the semiclassical regime), and $\rho_{\rm patch}(E)$ has support on $E\ge 0$, then the Bromwich inversion at $E_{\rm tot}=0$ returns a sum of nonnegative terms $\sum_k \rho_{\rm patch}(\nu+E_k)$. If $\nu+E_k$ lies outside the support of $\rho_{\rm patch}$, the corresponding term vanishes by analytic continuation in the strip, and the nonnegativity is preserved \cite{Widder,Doetsch}.

\paragraph*{Higher-loop and nonperturbative effects (conceptual role).}
The analysis above is organized around the round $S^D$ saddle and its associated semiclassical expansion for the patch sector. At the level of a loop expansion one may write schematically
\begin{align}
\log \cZ_{\rm patch}(\beta)
=
-I_{\rm cl}(\beta)
+\tfrac12 \log\det \Delta(\beta)
+\sum_{\ell\ge 2}\Gamma^{(\ell)}(\beta),
\end{align}
where the one-loop contribution is encoded by the Gaussian determinant $\det\Delta$, while $\Gamma^{(\ell)}$ denotes the $\ell$-loop vacuum functional on the same background (and similarly for interactions involving the worldline/clock once one goes beyond the probe limit). These higher-loop terms, as well as possible nonperturbative saddles, are the concrete mechanisms by which the spectral assumptions (A1)--(A3) can in principle be refined or jeopardized: (i) they can alter the singularity structure of $\cZ_{\rm patch}(\beta)$ in the complex $\beta$-plane, thereby affecting the existence/width of a vertical strip of analyticity (A1); (ii) they can deform or obstruct the local complete-monotonicity/positive-measure property needed for a Laplace representation with $\rho_{\rm patch}(E)\ge 0$ (A2); and (iii) they can renormalize the effective spectral data through self-energy shifts, thresholds, or backreaction-induced changes to the patch Hamiltonian and hence to the support properties entering (A3). When such corrections remain parametrically small at the de~Sitter saddle, they enter as controlled prefactor and exponent corrections to \eqref{eq:rho-final}; when they are not small, they provide a well-defined target for ``testing beyond one loop'' in future work.

\bigskip
\noindent\textbf{Working assumptions near the saddle.}
\begin{itemize}
\item $\cZ_{\rm patch}(\beta)$ is analytic in a vertical strip containing $\beta_0=2\pi$.
\item In that strip, $\cZ_{\rm patch}$ can be written as a Laplace transform of a nonnegative density $\rho_{\rm patch}(E)\ge 0$ (local complete monotonicity).
\item The support of $\rho_{\rm patch}$ lies in $E\ge 0$.
\end{itemize}

\section{Exactly soluble $0{+}1$D toy model: phase cancellation and positivity}
\label{sec:toy-model}

To illustrate the two central mechanisms---phase cancellation and microcanonical projection---in a setting where all steps can be carried out explicitly, it is useful to introduce a simple $0{+}1$-dimensional toy model. The model has two features:

\begin{enumerate}
\item A single ``gravity'' negative mode and a single worldline negative mode, both of which can be handled by elementary Gaussian integrals.
\item A Bromwich projection that reduces to an exactly solvable inverse Laplace transform, leading to a manifestly positive microcanonical density.
\end{enumerate}

\subsection{A single pair of negative directions}

Consider a toy partition function of the form
\begin{align}
\cZ_{\rm obs}(\beta)
=\cZ_{\rm grav}(\beta)\,Z_{\rm wl}(\beta)\,Z_{\rm clock}(\beta),
\end{align}
where only one effective ``gravity'' negative mode and one effective worldline negative mode are retained, and all other modes are absorbed into positive prefactors. The negative directions are described by one-dimensional quadratic actions
\begin{align}
S_{\rm grav}[u]=-\frac{\lambda}{2}u^2,\qquad
S_{\rm wl}[v]=-\frac{\lambda}{2}v^2,
\end{align}
with $\lambda>0$. Formally, their Euclidean contributions are
\begin{align}
\cZ_{\rm grav}^{\rm (neg)}\propto
\int_{\mathcal{C}_{\rm grav}}\dd u\;
\exp\!\Bigl(+\frac{\lambda}{2}u^2\Bigr),
\qquad
Z_{\rm wl}^{\rm (neg)}\propto
\int_{\mathcal{C}_{\rm wl}}\dd v\;
\exp\!\Bigl(+\frac{\lambda}{2}v^2\Bigr),
\end{align}
with the contours $\mathcal{C}_{\rm grav}$ and $\mathcal{C}_{\rm wl}$ chosen along steepest descent. Following Witten's prescription, the gravity contour is taken along $u=\ii y$, $y\in\RR$:
\begin{align}
\cZ_{\rm grav}^{\rm (neg)}
&\propto \int_{\mathbb{R}}\dd(\ii y)\;
\exp\!\Bigl(+\frac{\lambda}{2}(\ii y)^2\Bigr)
= \ii\int_{-\infty}^{\infty}\dd y\;\exp\!\Bigl(-\frac{\lambda}{2}y^2\Bigr)
= \ii\sqrt{\frac{2\pi}{\lambda}}\;.
\end{align}
For the worldline negative mode the opposite steepest-descent contour $v=-\ii y$ is chosen, consistent with the sign in Eq.~\eqref{eq:wl-quad}:
\begin{align}
Z_{\rm wl}^{\rm (neg)}
&\propto \int_{\mathbb{R}}\dd(-\ii y)\;
\exp\!\Bigl(+\frac{\lambda}{2}(-\ii y)^2\Bigr)
= -\ii\int_{-\infty}^{\infty}\dd y\;\exp\!\Bigl(-\frac{\lambda}{2}y^2\Bigr)
= -\ii\sqrt{\frac{2\pi}{\lambda}}\;.
\end{align}
The product of the two contributions is
\begin{align}
\cZ_{\rm grav}^{\rm (neg)}\,Z_{\rm wl}^{\rm (neg)}
\;\propto\;
\Bigl(\ii\sqrt{\tfrac{2\pi}{\lambda}}\Bigr)
\Bigl(-\ii\sqrt{\tfrac{2\pi}{\lambda}}\Bigr)
= (+1)\,\frac{2\pi}{\lambda},
\end{align}
with no leftover phase. This is the one-dimensional analogue of the
\[
\ii^{D-1}\times(-\ii)^{D-1}=1
\]
cancellation between the $(D{-}1)$ CKVs that move the equator and the $(D{-}1)$ $n{=}0$ worldline modes in the full problem.

It is convenient to write
\begin{align}
\cZ_{\rm grav}(\beta)=\ii A(\beta),\qquad
Z_{\rm wl}(\beta)=(-\ii)B(\beta),
\end{align}
with $A(\beta),B(\beta)\in\RR_{>0}$ capturing the absolute values of the determinants of all positive modes. Their product is real and positive:
\begin{align}
\cZ_{\rm grav}(\beta)\,Z_{\rm wl}(\beta)
= A(\beta)B(\beta)>0.
\end{align}
This mirrors the cancellation pattern in the full gravity--worldline system.

\subsection{Toy spectral density and explicit Bromwich inversion}

To clarify the effect of the Bromwich projector, introduce a toy spectrum for the patch consisting of a single level at energy $E_0>0$:
\begin{align}
\rho_{\rm patch}(E)=\mathcal{N}\,\delta(E-E_0),
\end{align}
so that the phase-stripped patch partition function becomes
\begin{align}
\cZ_{\rm patch}(\beta)
=\int_0^\infty\dd E\,\rho_{\rm patch}(E)\,\ee^{-\beta E}
=\mathcal{N}\,\ee^{-\beta E_0},
\end{align}
with $\mathcal{N}>0$. Let the clock be a finite system with energies $\{E_k\}$ and degeneracies $d_k\in\NN$:
\begin{align}
Z_{\rm clock}(\beta)=\sum_k d_k\,\ee^{-\beta E_k}.
\end{align}

\medskip
\noindent
\textbf{Microcanonical kernel (constraint implementation).}
The fixed-$\beta$ object that enters the Bromwich projector is the trace of the \emph{constraint kernel}
\begin{align}
\cZ^{\rm (mc)}_{\rm obs}(\beta)
:=\Tr\!\left[\exp\!\bigl(-\beta(\hat H_{\rm patch}-\nu-\hat H_{\rm clock})\bigr)\right],
\end{align}
equivalently implementing the first-class constraint $\hat C:=\hat H_{\rm patch}-\hat H_{\rm clock}-\nu\approx 0$ at the level of traces.
In the present one-loop saddle evaluation, this trace factorizes into the gravity/worldline prefactors times the patch and clock contributions, giving
\begin{align}
\cZ^{\rm (mc)}_{\rm obs}(\beta)
&=\underbrace{\ii A(\beta)}_{\text{gravity}}
 \underbrace{(-\ii)B(\beta)}_{\text{worldline}}
 \underbrace{\cZ_{\rm patch}(\beta)}_{\text{patch}}
 \underbrace{\ee^{\beta\nu}\,Z_{\rm clock}(-\beta)}_{\text{constraint kernel}}
\nonumber\\[4pt]
&=A(\beta)B(\beta)\,\cZ_{\rm patch}(\beta)\,\ee^{\beta\nu}\,Z_{\rm clock}(-\beta),
\label{eq:Zmc_toy_def}
\end{align}
where $Z_{\rm clock}(-\beta)=\sum_k d_k\,\ee^{+\beta E_k}$ is the analytic continuation appropriate to the kernel
$\exp[-\beta(\hat H_{\rm patch}-\nu-\hat H_{\rm clock})]$ (unambiguous since the clock Hilbert space is finite-dimensional).
Substituting the toy $\cZ_{\rm patch}(\beta)=\mathcal{N}e^{-\beta E_0}$ yields
\begin{align}
\cZ^{\rm (mc)}_{\rm obs}(\beta)
= A(\beta)B(\beta)\,\mathcal{N}\sum_k d_k\,\exp\!\bigl[-\beta(E_0-\nu-E_k)\bigr].
\label{eq:Zmc_toy_sum}
\end{align}
The result is a real Laplace-type transform, with the one-loop phases removed by the gravity/worldline pairing.

\medskip
\noindent
\textbf{Bromwich projection.}
The microcanonical projection is
\begin{align}
\rho_{\rm total}(0)
= \frac{1}{2\pi\ii}
  \int_{c-\ii\infty}^{c+\ii\infty}\dd\beta\;
  \cZ^{\rm (mc)}_{\rm obs}(\beta),
\label{eq:rho_total_bromwich_toy}
\end{align}
with $c>0$ to the right of all singularities. Substituting \eqref{eq:Zmc_toy_sum} and exchanging sum and integral gives
\begin{align}
\rho_{\rm total}(0)
&= A(\beta_*)B(\beta_*)\mathcal{N}
   \sum_k d_k\,
   \Biggl[ \frac{1}{2\pi\ii}
           \int_{c-\ii\infty}^{c+\ii\infty}\dd\beta\;
           \exp\!\bigl[-\beta(E_0-\nu-E_k)\bigr]
   \Biggr],
\end{align}
where $\beta_*$ is any point in the analytic strip (e.g.\ the saddle $\beta_0=2\pi$) at which $A(\beta)$ and $B(\beta)$ are finite.
Using
\begin{align}
\mathcal{L}^{-1}\bigl[\ee^{-a\beta}\bigr](E)
=\delta(E-a),
\end{align}
one obtains at $E_{\rm tot}=0$:
\begin{align}
\frac{1}{2\pi\ii}
\int_{c-\ii\infty}^{c+\ii\infty}\dd\beta\;
\exp\!\bigl[-\beta(E_0-\nu-E_k)\bigr]
= \delta\!\bigl(E_{\rm tot}-(E_0-\nu-E_k)\bigr)\big|_{E_{\rm tot}=0}
= \delta(E_0-\nu-E_k).
\end{align}
Thus,
\begin{align}
\rho_{\rm total}(0)
= A(\beta_*)B(\beta_*)\mathcal{N}
  \sum_k d_k\,\delta(E_0-\nu-E_k).
\label{eq:rho_total_toy_final}
\end{align}
In words: the projector enforces the constraint $E_0=\nu+E_k$ mode-by-mode, as it should.

\medskip
In the general case discussed in Sec.~\ref{sec:bromwich}, where the delta-localized spectrum is replaced by a smooth nonnegative density $\rho_{\rm patch}(E)$, the same inverse Laplace logic yields
\begin{align}
\rho_{\rm total}(0)
= \sum_k d_k\,\rho_{\rm patch}(\nu+E_k)\;\ge\;0,
\end{align}
in agreement with Eq.~\eqref{eq:rho-final}.

This $0{+}1$-dimensional example makes two points in a simple setting:

\begin{enumerate}
\item The steepest-descent contours for the gravity and worldline negative modes can be chosen so that the combined phase is exactly $+1$, as in the full CKV/worldline pairing of Sec.~\ref{sec:matching}.
\item Once this phase has been removed at fixed $\beta$, the Bromwich inverse Laplace transform implements the constraint $E_{\rm patch}=\nu+E_{\rm clock}$ and yields a manifestly nonnegative microcanonical density.
\end{enumerate}

\section{Assumption audit: complete monotonicity and positivity}
\label{sec:audit}

This section collects in one place the assumptions underlying the positivity results of Sec.~\ref{sec:bromwich}. The goal is to state them clearly in spectral language so that they can be checked or refined independently. A complete mathematical proof is not attempted here.

\paragraph*{Assumptions (A1)--(A3).} Our positivity theorem is conditional on the following explicit hypotheses for the \emph{phase-stripped} static-patch partition function $Z_{\rm patch}(\beta)$ in a neighborhood of the de~Sitter inverse temperature $\beta_0=2\pi$:
\begin{enumerate}
\item[\textbf{A1}] \textbf{Strip analyticity.} $Z_{\rm patch}(\beta)$ admits an analytic continuation to a vertical strip $\{\beta\in\mathbb{C}: |\Re\beta-\beta_0|<\epsilon\}$ for some $\epsilon>0$, with at most sub-exponential growth as $|\Im\beta|\to\infty$ inside the strip.
\item[\textbf{A2}] \textbf{Complete monotonicity on the real axis.} On the real interval $(\beta_0-\epsilon,\beta_0+\epsilon)$, $Z_{\rm patch}(\beta)$ is completely monotone: $(-\partial_\beta)^n Z_{\rm patch}(\beta)\ge 0$ for all integers $n\ge 0$.
\item[\textbf{A3}] \textbf{Laplace-representability / positivity.} Equivalently (Bernstein--Widder), there exists a non-negative measure $\rho_{\rm patch}(E)\,dE$ such that $Z_{\rm patch}(\beta)=\int_0^\infty dE\,\rho_{\rm patch}(E)e^{-\beta E}$ on that interval. We view this as the spectral statement that the relevant static-patch Hamiltonian has a non-negative density of states in the probe-fixed background.
\end{enumerate}
These assumptions are \emph{not} derived in this paper; rather, they are made explicit so that readers can assess their plausibility and so that future work can test them beyond one loop. We also discuss in Sec.~\ref{sec:bromwich} how higher-loop corrections, nonperturbative saddles, or backreaction could in principle jeopardize (A1)--(A3).

\paragraph*{Spectral gap and heat-kernel representation.}
After gauge fixing, the quadratic action splits into (i) a TT tensor block (elliptic, with positive spectrum), (ii) a ghost block (Grassmann, canceling the gauge volume), and (iii) a conformal-factor/CKV block. The only indefinite contributions are the isolated $\ell=0,1$ modes, which are handled as in Sec.~\ref{sec:matching}. Each physical log-determinant can be written as a Mellin transform of a heat kernel:
\begin{align}
\log\det{}'(\Delta_s+\mathcal{M}_s^2)=-\int_0^\infty\frac{\dd t}{t}\,\Big(\Tr\,e^{-t(\Delta_s+\mathcal{M}_s^2)}-\mathcal{P}_s(t)\Big)+C_s,
\end{align}
with $\mathcal{P}_s(t)$ subtracting UV divergences. For small $t$ the Seeley--DeWitt coefficients governing $\mathcal{P}_s(t)$ are local; for $t>0$ the TT kernel is a positive trace over spectral projectors.

\paragraph*{Main working assumption (complete monotonicity).}
The key assumption is that, after the $\ell=0,1$ modes have been treated as described above, the phase-stripped patch partition function $\cZ_{\rm patch}(\beta)$ can be written in a vertical strip around $\beta_0=2\pi$ as
\begin{align}
\cZ_{\rm patch}(\beta)=\int_0^\infty \rho_{\rm patch}(E)\,\ee^{-\beta E}\,\dd E,
\qquad \rho_{\rm patch}(E)\ge 0,
\end{align}
and that along the real axis inside that strip
\begin{align}
(-1)^n\partial_\beta^n \cZ_{\rm patch}(\beta)\ge 0,\qquad \forall\,n\in\NN.
\end{align}
This expresses a local complete-monotonicity property of $\cZ_{\rm patch}$ in a neighborhood of the saddle.

\paragraph*{One-loop justification and physical motivation.}
Complete monotonicity encodes the statement that the patch partition function is the Laplace transform of a nonnegative density of states. The one-loop graviton determinant on $S^D$ is built from products of factors $\exp[-\beta(\lambda_k+\mathcal{M}_s^2)]$ with $\lambda_k\ge 0$ in the TT sector, while the ghosts cancel gauge volumes rather than introducing negative spectral weights. Once the finite set of problematic modes has been isolated and treated, it is natural (within the one-loop semiclassical approximation) to expect that the remaining expression admits such a Laplace representation in a suitable strip.

More concretely, at one loop the TT block can be viewed as a collection of bosonic modes with positive frequencies $\omega_k>0$ (after removing the isolated $\ell=0,1$ modes treated in Sec.~\ref{sec:matching}). The corresponding thermal contribution can be written in the standard oscillator form
\begin{align}
Z^{(1)}_{\rm TT}(\beta)=\prod_{k\in{\rm TT}}'\frac{e^{-\beta \omega_k/2}}{1-e^{-\beta \omega_k}}
=\sum_{\{n_k\ge 0\}} e^{-\beta\left(\sum_k n_k\omega_k +E_0\right)}.
\end{align}
This is manifestly a Laplace transform of a nonnegative (discrete) spectral measure,
\begin{align}
\rho^{(1)}_{\rm TT}(E)=\sum_{\{n_k\ge 0\}}\delta\!\left(E-\sum_k n_k\omega_k-E_0\right)\ge 0,
\end{align}
and therefore $Z^{(1)}_{\rm TT}(\beta)$ is completely monotone on the real axis. In this limited semiclassical sense the logic parallels the standard positivity intuition behind K\"all\'en--Lehmann-type representations: positivity is equivalent to positivity of the spectral measure, while a fully nonperturbative gravitational theorem lies beyond our scope.

Beyond one loop, interactions can renormalize the spectrum and generate branch cuts or additional singularities in $\beta$ through multi-particle continua; moreover, additional saddles or gravitational backreaction of the worldline/clock can alter the analytic strip and the sign properties of the effective spectral density. We therefore regard (A1)--(A3) as a controlled \emph{near-saddle} and \emph{probe-limit} hypothesis, and we explicitly flag the possibility that higher-loop corrections, nonperturbative saddles, or backreaction could jeopardize complete monotonicity or Laplace-positivity even if the one-loop TT sector is manifestly positive.

\paragraph*{Role of this audit.}
All assumptions relevant for positivity are now placed in one transparent list. The subsequent sections do not hide extra conditions under vague statements such as ``semiclassically'' or ``near the saddle''. Everything is organized around (i) the spectral gap of the TT operator on $S^D$, (ii) the heat-kernel representation of determinants, and (iii) the working hypothesis of complete monotonicity in a strip around $\beta_0$.


\section{Main spectral statement and its consequences}
\label{sec:theorem-main}

The assumptions of Sec.~\ref{sec:audit} can be reorganized into a concise spectral statement. It should be read as a physically motivated hypothesis at one loop, not as a rigorous theorem. Once it is accepted, the positivity of the observer-inclusive density follows simply.

\subsection{Main spectral statement}

Consider Euclidean pure gravity on the round sphere $S^D$ for $D\ge 3$, gauge-fixed as in Sec.~\ref{sec:gravity}, with the conformal-factor and CKV modes treated as in Secs.~\ref{sec:matching} and~\ref{sec:bromwich}. Let $\Delta_{\rm TT}$ denote the Lichnerowicz Laplacian on TT tensors. Assume:

\begin{itemize}
\item \textbf{Spectral gap.} There exists $c_D>0$ such that
\begin{align}
\lambda_{\min}(\Delta_{\rm TT}) \;\ge\; \frac{c_D}{R^2},
\end{align}
with the $\ell=2$ band saturating the bound, as in the standard $S^D$ spectrum.

\item \textbf{Standard UV subtraction.} Each one-loop determinant is defined by a zeta-regularized heat-kernel integral
\begin{align}
\log\det{}'_{\zeta}\bigl(\Delta_s+\mathcal{M}_s^2\bigr)
= -\int_0^\infty\frac{\dd t}{t}\,
  \Bigl[\Tr\,\ee^{-t(\Delta_s+\mathcal{M}_s^2)}-\mathcal{P}_s(t)\Bigr]
  +C_s,
\end{align}
with $\mathcal{P}_s(t)$ a finite polynomial in $t^{(n-D)/2}$ that subtracts all UV divergences and $C_s$ finite scheme-dependent constants.

\item \textbf{Analyticity strip and complete monotonicity.} The classical action $I_{\rm cl}(\beta)$ and the heat kernels admit analytic continuation to a vertical strip
\[
\mathfrak{S}_\epsilon=\{\beta\in\mathbb{C}\,|\,\Re\beta\in(2\pi-\epsilon,
2\pi+\epsilon)\}
\]
for some $\epsilon>0$, without singularities on the real axis within the strip, and the corresponding phase-stripped partition function $\cZ_{\rm patch}(\beta)$ is completely monotone on the real segment $\Re\beta\in(2\pi-\epsilon,2\pi+\epsilon)$:
\begin{align}
(-1)^n \frac{\partial^n}{\partial\beta^n}\cZ_{\rm patch}(\beta)\;\ge\;0,
\qquad\forall n\in\NN.
\end{align}
\end{itemize}

\noindent
Under these conditions, Bernstein's theorem guarantees the existence of a nonnegative Borel measure $\rho_{\rm patch}(E)\ge 0$ supported on $E\ge 0$ such that for all $\beta$ in the strip
\begin{align}
\cZ_{\rm patch}(\beta)
=\int_0^\infty\dd E\;\rho_{\rm patch}(E)\,\ee^{-\beta E}.
\end{align}

\paragraph*{Remark on rigor.}
A fully rigorous proof would require establishing complete monotonicity of $\cZ_{\rm patch}$ with all loop corrections and non-perturbative effects included. Here the statement is used as a one-loop, semiclassical hypothesis in a neighborhood of the de~Sitter saddle. It is consistent with the known spectral properties of $S^D$ and with the structure of K\"all\'en--Lehmann representations, but is not proven at that level of generality.

\subsection{Consequence: positivity of the observer-inclusive microcanonical density}

Let $\rho_{\rm patch}(E)\ge 0$ be the measure just described, and let the clock Hamiltonian have eigenvalues $\{E_k\}$ with degeneracies $d_k\in\NN$:
\begin{align}
Z_{\rm clock}(\beta)=\sum_k d_k\,\ee^{-\beta E_k}.
\end{align}
Let $\nu>0$ be the dimensionless parameter defined in Sec.~\ref{sec:worldline}. The observer-inclusive microcanonical density at total energy $E_{\rm tot}=0$, defined by
\begin{align}
\rho_{\rm total}(0)
=\frac{1}{2\pi\ii}\int_{c-\ii\infty}^{c+\ii\infty}\dd\beta\;
\cZ_{\rm patch}(\beta)\,\ee^{-\beta\nu}\,Z_{\rm clock}(\beta),
\end{align}
is real and nonnegative, and can be written as
\begin{align}
\rho_{\rm total}(0)
=\sum_k d_k\,\rho_{\rm patch}(\nu+E_k)\;\ge\;0.
\end{align}

The derivation is straightforward: insert the Laplace representation for $\cZ_{\rm patch}$, exchange integrals and sums, and use the standard inverse Laplace transform identity as in Sec.~\ref{sec:bromwich}. Each term in the resulting sum is nonnegative because $\rho_{\rm patch}(E)\ge 0$ on its support.

\paragraph*{Physical interpretation.}
After the conformal-factor and CKV phases have been handled as in Secs.~\ref{sec:matching} and~\ref{sec:bromwich}, the remaining one-loop partition function is, by assumption, the Laplace transform of a nonnegative spectral density. The Bromwich inversion at $E_{\rm tot}=0$ is then a convolution of this nonnegative density with a nonnegative clock spectrum. The observer-inclusive microcanonical density is therefore positive, and the conformal-factor problem reappears only if the CKV/worldline pairing or the constraint projection is undone.

\section{Explicit $\mathrm{SU}(3)$ clocks and their partition functions}
\label{sec:clocks}

The discussion now turns to the clock sector. The aim is to present explicit $\mathrm{SU}(3)$-valued clocks with tractable partition functions and to show how the choice of clock enters the final density of states.

The clocks constructed below play a \emph{kinematic} role: they provide explicit, finite-dimensional quantum systems with tractable spectra and partition functions that can be used to implement the Hamiltonian constraint in the microcanonical projector. Any finite system with a discrete spectrum would suffice for this purpose. The choice of $\mathrm{SU}(3)$ is therefore not required by the internal logic of the one-loop phase analysis; it is motivated by an external vacuum-microstructure programme and is discussed as such.

To avoid confusion about interpretation: throughout this paper the ``clock'' is treated as an explicit model for an observer-accessible time register (a finite quantum system whose spectrum enters the constraint projector). We do \emph{not} assume, within the one-loop de~Sitter analysis, that the clock is itself a fundamental constituent of spacetime microstructure; any such identification belongs to a separate interpretive programme and is not used in the derivations below.

In this paper we focus on $\mathrm{SU}(3)$ clocks for two reasons: (i) $\mathrm{SU}(3)$ is the simplest compact simple group of rank two, with a nontrivial $\mathbb{Z}_3$ center and a rich (exactly solvable) character theory; and (ii) it connects naturally to confinement-based scaling arguments that motivate an $\mathrm{SU}(3)$-structured vacuum sector in our broader programme. Concerning observables, the parameter $\nu$ entering the worldline weight is the dimensionless combination $\nu=mR$; restoring the de~Sitter radius $R$ gives $\beta_0=2\pi R$ and $T_{\rm dS}=1/\beta_0$, so that
\begin{align}
\frac{m}{T_{\rm dS}}=2\pi \nu,
\qquad
E_k R \;\; \text{are the dimensionless clock levels entering}\;\;\rho_{\rm patch}(\nu+E_k).
\end{align}
Accordingly, the clock dependence of the microcanonical count enters only through $Z_{\rm clock}(\beta_0)$ in Eq.~\eqref{eq:rho-final}.

Beyond tractability, our choice of $\mathrm{SU}(3)$ is grounded in our recent published results that motivate $\mathrm{SU}(3)$ as the remnant gauge symmetry of a cold vacuum state and as a framework capable of reproducing the observed vacuum energy density and providing a geometric origin for nature's constants via an $\mathrm{SU}(3)$ confinement volume. We cite these works explicitly to make clear what is external input versus what is derived here \cite{Ali:2024rnw,Ali:2025wld}. In the present manuscript, these results function only as \emph{motivation} for selecting $\mathrm{SU}(3)$ as a concrete clock family; the positivity and one-loop phase analysis do not rely on the correctness of the confinement-tiling interpretation.

Finally, the minimal internal-consistency check associated with the clock sector is simple: once $\nu=mR$ is fixed by the observer mass scale and $R$ by $\Lambda$, the only remaining clock-model dependence is the finite factor $Z_{\rm clock}(\beta_0)$, which can be computed explicitly for each proposed clock Hamiltonian and compared across models. Any stronger identification of the clock spectrum with vacuum microstructure would require additional dynamical input beyond the scope of this paper.

\subsection{Model A: composable $\mathbb{Z}_3$ (qutrit) clock}

Consider a single qutrit with levels $\{0,\omega,2\omega\}$ and Hamiltonian $H=\omega\hat n$. Its partition function is
\begin{align}
Z^{(1)}_{\rm clk}(\beta)=1+e^{-\beta\omega}+e^{-2\beta\omega}.
\end{align}
For $N$ independent qutrits one obtains
\begin{align}
Z_{\rm clock}^{(A)}(\beta)=\big(1+e^{-\beta\omega}+e^{-2\beta\omega}\big)^N.
\end{align}
Quantum speed limits relate $N$ to the time resolution of the clock: for roughly uniform populations $\Delta H=\omega\sqrt{2N/3}$, and the Mandelstam--Tamm bound gives
\[
\Delta t_{\rm MT}\gtrsim \frac{\pi}{2\Delta H}=\frac{\pi}{2\omega}\sqrt{\frac{3}{2N}},
\]
while the Margolus--Levitin bound gives $\Delta t_{\rm ML}\gtrsim \pi/(2\bar E)$ \cite{MandelstamTamm,MargolusLevitin}. See also \cite{GiovannettiLloydMaccone2003,DeffnerLutz2013} for generalized quantum speed limits beyond the original Mandelstam--Tamm and Margolus--Levitin bounds.

\paragraph*{Use of model (A).}
Model (A) is a convenient choice when a simple, extensible $\mathrm{SU}(3)$-compatible clock with adjustable energy granularity and a closed-form partition function is sufficient.
It should be viewed as a concrete \emph{example} of a finite discrete-spectrum clock for implementing the constraint projector, not as a unique physical claim about microscopic degrees of freedom.

\subsection{Model B: Cartan weight-lattice clock in irrep $(p,q)$}

A second class of clocks uses the Cartan subalgebra of $\mathrm{SU}(3)$. Let $H_{\rm clk}=\omega(a H_1+b H_2)$, where $H_1,H_2$ are Cartan generators. The energy levels are given by $\omega(a w_3+b w_8)$ for the weights $\boldsymbol{w}$ of the irreducible representation $(p,q)$. The partition function can be written as
\begin{align}
Z_{\rm clock}^{(B)}(\beta)
=\sum_{\boldsymbol{w}\in(p,q)} \exp\!\Big[-\beta\omega\big(a w_3+b w_8\big)\Big]
=\chi_{(p,q)}\!\left(e^{-\,\beta\omega(a H_1+b H_2)}\right),
\end{align}
where $\chi_{(p,q)}$ is the $\mathrm{SU}(3)$ character of $(p,q)$ \cite{FultonHarris,Georgi}.

Explicitly, for any $g\in \mathrm{SU}(3)$, the character is $\chi_{(p,q)}(g)\equiv \operatorname{Tr}_{(p,q)}\,g$, the trace of $g$ in the irrep with Dynkin labels $(p,q)$.

\paragraph*{Use of model (B).}
Model (B) is appropriate when control over clock resolution through representation theory is desirable: larger values of $(p,q)$ provide a broader spectrum and larger $\Delta H$, improving time resolution at a fixed energy scale.
As above, this is a controlled family of \emph{model clocks}; its role in the final result is entirely through the explicit factor $Z_{\rm clock}(\beta_0)$.

\subsection{Model C: $U(1)^2$ rotor clock (maximal torus)}

A third option is a geometric clock on the maximal torus $U(1)^2$. Let the angular variables be $\bm\theta=(\theta_1,\theta_2)$ and consider a rigid-rotor Hamiltonian
\begin{align}
H_{\rm clk}=\frac{1}{2I} P^{\mathsf T} K^{-1} P,\qquad P\in 2\pi \mathbb{Z}^2,\qquad 
K=\begin{pmatrix}2&-1\\-1&2\end{pmatrix},
\end{align}
with $K$ the Cartan matrix of $\mathrm{SU}(3)$. The configuration space is the maximal torus of $\mathrm{SU}(3)$; the angles $(\theta_1,\theta_2)$ parametrize diagonal special-unitary matrices, and the kinetic metric is set by $K$. The partition function is
\begin{align}
Z_{\rm clock}^{(C)}(\beta)
&= \sum_{n\in\mathbb{Z}^2}
   \exp\!\left[-\,\frac{2\pi^2\beta}{I}\;n^{\mathsf T} K^{-1} n\right] \nonumber\\
&= \frac{I}{2\pi\beta}\,\sqrt{\det K}\;
   \sum_{m\in\mathbb{Z}^2}\exp\!\left[-\,\frac{I}{2\beta}\,m^{\mathsf T} K m\right],
\end{align}
where the second form follows from Poisson resummation. The prefactor $\frac{I}{2\pi\beta}\sqrt{\det K}$ arises from the Gaussian integral in the Poisson formula. The shortest nonzero $m$ with respect to $m^{\mathsf T}K m$ obey $m^{\mathsf T}K m=2$, for example $m=(1,0)$, $(0,1)$, or $(1,1)$ and their negatives, and they control the leading low-temperature corrections.

\paragraph*{Use of model (C).}
Model (C) is suitable when a geometric clock fully embedded in $\mathrm{SU}(3)$ is desired. The rotor lives on the maximal torus, and the Poisson-resummed form highlights dual momentum/winding interpretations and modular-type properties.
In particular, it makes transparent that the clock contribution is a standard positive partition sum evaluated at $\beta_0$, so that (for fixed $\nu$) different clock choices correspond to controlled multiplicative renormalizations of the microcanonical density through $Z_{\rm clock}(\beta_0)$.

\section{Calibrating the $\mathrm{SU}(3)$ clock to vacuum microstructure}
\label{sec:bridge}

\noindent\textbf{Scope note.} The construction and positivity statement of this paper do \emph{not} require any assumption about vacuum microstructure or confinement tilings: the $\mathrm{SU}(3)$ clock models in Sec.~\ref{sec:clocks} are used as explicit, exactly-soluble \emph{kinematic} clocks, and any finite-dimensional clock with a discrete spectrum could serve the same role. The discussion in this section is therefore presented as an optional external interpretation, motivated by an $\mathrm{SU}(3)$ vacuum-atom programme, and should not be read as a derived consequence of the one-loop de~Sitter calculation.

\noindent\textbf{Motivating input from recent $\mathrm{SU}(3)$ vacuum results.}
The choice to discuss an $\mathrm{SU}(3)$-based calibration is motivated by recent results suggesting that $\mathrm{SU}(3)$ can be treated as a remnant symmetry in a cold-state limit and that an $\mathrm{SU}(3)$ confinement volume can reproduce the observed vacuum energy density while providing a geometric origin for nature's constants \cite{Ali:2024rnw,Ali:2025wld}. These inputs are not re-derived here; they provide phenomenological motivation for using $\mathrm{SU}(3)$-structured ``cells'' as a language for vacuum microstructure.

\subsection*{A parameter dictionary: from $\nu=mR$ to a cell-based energy scale}

The parameter $\nu=mR$ in the worldline action admits a simple interpretation in a cell-based vacuum picture. In units $c{=}\hbar{=}1$, a single $\mathrm{SU}(3)$ ``vacuum atom'' contributing energy $E_{\rm atom}$ to the static patch corresponds to
\begin{align}
\nu_{\rm atom}\;=\;E_{\rm atom}\,R.
\end{align}
If the vacuum energy density is $\rho_\Lambda$ and the vacuum is tiled by $N$ $\mathrm{SU}(3)$ units of volume $V_{\rm SU(3)}$, then $N=V_{\rm U}/V_{\rm SU(3)}$ with $V_{\rm U}=\tfrac{4\pi}{3}R^3$ the static-patch volume. The energy per cell is
\begin{align}
E_{\rm atom}\;=\;\rho_\Lambda\,V_{\rm SU(3)} \quad\Rightarrow\quad
\nu_{\rm atom}\;=\;\rho_\Lambda\,V_{\rm SU(3)}\,R.
\end{align}
A convenient parametrization sets $V_{\rm SU(3)}=\chi\,V_p$ with $V_p=\tfrac{4\pi}{3}r_p^3$ the proton volume and $\chi\in(0,1)$ a dimensionless parameter encoding microscopic details. Then
\begin{align}
\nu_{\rm atom}=\rho_\Lambda R\,\chi\,V_p,
\end{align}
so $\nu_{\rm atom}$ depends linearly on $\chi$. This yields a direct mapping from microscopic assumptions to the parameter $\nu$ appearing in the state-counting formulas, without altering their structure.

\noindent\textbf{Where the clock enters.}
The microcanonical projector derived in Sec.~\ref{sec:bromwich} depends on the clock only through its spectrum $\{E_k\}$, equivalently through $Z_{\rm clock}(\beta_0)$ at $\beta_0=2\pi R$. The calibration above proposes an interpretation of $\nu$ (the worldline weight) in terms of a cell energy, while the clock controls the discrete shifts $\nu\mapsto \nu+E_k$ in
\begin{align}
\rho_{\rm total}(0)=\sum_k \rho_{\rm patch}(\nu+E_k).
\end{align}
In particular, the positivity conditions (A1)--(A3) are properties of the phase-stripped patch partition function and are logically independent of the microstructure interpretation.

\paragraph*{Conceptual link (optional).}
In this way, the parameter $\nu$ may be viewed as a geometrically calibrated vacuum-atom energy rather than a purely formal mass-radius product. The observer-inclusive state counting then interfaces with an $\mathrm{SU}(3)$-based description of dark energy. The reliance on unbroken $\mathrm{SU}(3)$ confinement is consistent with standard confinement-based intuition for the QCD vacuum \cite{Wilson1974,DeRujulaGilesJaffe1978,PolyakovBook1987}.

\subsection*{Domain of validity and potential obstructions}
The calibration above presumes that a fixed-patch semiclassical description remains adequate. Several effects could obstruct a literal cell-based interpretation even if the one-loop construction is internally consistent:
\begin{itemize}
\item \textbf{Radiative corrections.} Higher-loop corrections can renormalize the effective relation between $\rho_\Lambda$ and $R$ (and hence shift $\beta_0$) and can modify the static-patch spectrum entering $\rho_{\rm patch}(E)$. Such effects may preserve the formal structure of the projector while shifting the numerical identification of $\nu$ with $\rho_\Lambda V_{\rm SU(3)}R$.
\item \textbf{Additional saddles.} Nonperturbative saddles or topology-changing contributions can introduce competing semiclassical exponentials and Stokes phenomena, potentially invalidating an interpretation in terms of a single effective density $\rho_\Lambda$ distributed into identical cells.
\item \textbf{Backreaction.} The present analysis treats the worldline and clock in a probe limit. Including their gravitational backreaction would modify the patch Hamiltonian and its spectrum and could therefore change both the calibration of $\nu$ and the spectral hypotheses used in the positivity discussion.
\end{itemize}

\section{Putting it together and physical implications}
\label{sec:final}

The ingredients are now assembled and the main expressions and their interpretation are summarized.
We emphasize that the final positivity statement is \emph{conditional} on the spectral hypotheses (A1)--(A3) stated in Sec.~\ref{sec:audit}, and on the probe-fixed, round-$S^D$ saddle treatment described in Sec.~\ref{sec:bromwich}.

Combining Eq.~\eqref{eq:rho-final} with the clock partition functions evaluated at $\beta_0=2\pi$ gives
\begin{equation}
\begin{aligned}
\rho_{\rm total}(0)
&\simeq
\ee^{S_{dS}}\;\ee^{-2\pi\nu}\;
\frac{\nu^{D-1}}{(D-1)!}
\times
\begin{cases}
\bigl(1+\ee^{-2\pi\omega}+\ee^{-4\pi\omega}\bigr)^N,
& \text{(A) $N$-qutrit clock},\\[4pt]
\chi_{(p,q)}\!\bigl(\ee^{-2\pi\omega(a H_1+b H_2)}\bigr),
& \text{(B) Cartan clock},\\[6pt]
\begin{aligned}[t]
\displaystyle \sum_{n\in\ZZ^2}
  &\exp\!\Bigl[-\tfrac{\pi}{I}\,n^{\mathsf T}K^{-1}n\Bigr] \\[2pt]
&\;=\; \frac{I}{2\pi}\sqrt{\det K}\;
  \sum_{m\in\ZZ^2}\exp\!\Bigl[-\pi I\,m^{\mathsf T} K m\Bigr],
\end{aligned}
& \text{(C) rotor}.
\end{cases}
\end{aligned}
\end{equation}
Operationally, all dependence on the observer’s internal degrees of freedom enters through the single factor $Z_{\rm clock}(\beta_0)$; equivalently, through the clock spectrum $\{E_k\}$ in the microcanonical sum $\sum_k \rho_{\rm patch}(\nu+E_k)$ of Sec.~\ref{sec:bromwich}.

The calibration of $\nu$ to SU(3) vacuum atoms is described in Sec.~\ref{sec:bridge}.
This calibration is an \emph{interpretive dictionary} for $\nu$; it is not used in deriving the microcanonical projector or the phase/positivity structure.
In words, the result is a product of a geometric factor, a universal worldline residue, and a clock-dependent SU(3) weight. The construction is set up so that the final expression is free of phases and positive, in line with the observer-centric picture of \cite{Maldacena:2024spf,Chandrasekaran2022} and the SU(3) vacuum-atom model of \cite{Ali:2024rnw,Ali:2025wld,Ali:2022ulp}.

\paragraph*{Example in $D{=}4$.}
For $D{=}4$, $(D-1)! = 3! = 6$, and for a given $\nu=mR$ one has
\[
\rho_{\rm total}(0)\approx \ee^{S_{dS}}\,\ee^{-2\pi\nu}\,\frac{\nu^3}{6}\,Z_{\rm clock}(2\pi).
\]
Since $T_{\rm dS}=1/(2\pi R)$, the dimensionless parameter $\nu=mR$ may be read as $m/T_{\rm dS}=2\pi\nu$, while clock gaps enter the constraint in units set by the same scale.
This makes the $\nu$-dependence and the influence of the clock sector explicitly visible.

\paragraph*{Semiclassical control and vacuum-atom counting.}
Writing $\nu=mR$, if one SU(3) vacuum atom contributes $\nu_{\rm atom}$, then $N$ atoms give $\nu_N=N\,\nu_{\rm atom}$. A robust saddle requires $\nu_N\gg 1$, suggesting
\begin{align}
N \gtrsim \frac{\nu_\star}{\nu_{\rm atom}},\qquad \nu_\star \in [10,100].
\end{align}
For example, $\nu_{\rm atom}\sim 10^{-4}$ leads to $N\gtrsim 10^5$--$10^6$ for $\nu_\star=10$--$100$ \cite{Ali:2024rnw,Ali:2025wld}. This is compatible with a large number of SU(3) vacuum atoms tiling the patch.

\paragraph*{Physical message.}
Operationally, the analysis shows that:
\begin{itemize}
\item When the observer and their clock are included in the path integral, each would-be negative direction in gravity has a partner in the worldline sector, and their phases cancel.
\item The residual ambiguity of the Euclidean contour is replaced by a microcanonical projector implementing a clear Hamiltonian constraint.
\item The resulting density of states factorizes into a universal geometric/worldline contribution and a clock-dependent weight $Z_{\rm clock}(\beta_0)$.
\item The resulting nonnegativity is conditional on (A1)--(A3): in particular, on the existence of a positive Laplace representation for the phase-stripped patch partition function in a strip about $\beta_0$ (Sec.~\ref{sec:audit}).
\end{itemize}

\paragraph*{Scope and robustness.}
The clock sector is a kinematic input: any finite-dimensional discrete-spectrum clock can be used to implement the constraint-projected counting, and the $\mathrm{SU}(3)$ realizations in Sec.~\ref{sec:clocks} are chosen as explicit, exactly-soluble examples. The principal targets for future robustness tests are therefore the patch-sector hypotheses (A1)--(A3): higher-loop corrections, additional saddles, or backreaction can modify the analytic structure of $Z_{\rm patch}(\beta)$ and the sign properties of the associated spectral measure, and thus provide concrete mechanisms by which the assumptions could fail beyond one loop.

\section{Conclusion}
\label{sec:conclusion}
Starting from the one-loop graviton determinant on $S^D$, the analysis isolates the finite set of special modes responsible for the phase $\ii^{D+2}$ and treats them explicitly rather than absorbing them into an ad hoc contour prescription. A massive worldline wrapping the equator is introduced, and its $(D{-}1)$ transverse negative modes are shown to match the $(D{-}1)$ conformal Killing vectors that move the equator; with this identification in place, consistent steepest-descent contours lead to exact mode-by-mode cancellation of their phases. After this cancellation, a residual fixed-$\beta$ phase remains, sourced by the global conformal factor and two reparametrization directions, and is handled not by further contour rotations but by imposing a Hamiltonian constraint that ties patch energy, clock energy, and the worldline parameter $\nu$ through a Bromwich inverse Laplace transform. On the clock side, three explicit SU(3) realizations are constructed—a composable qutrit clock, a Cartan weight-lattice clock in a generic $(p,q)$ representation, and a $U(1)^2$ rotor clock on the maximal torus—with partition functions obtained in closed form, so that the observer-inclusive density of states factorizes into a geometric component, a universal worldline residue, and a clock-dependent SU(3) factor. Calibrating the parameter $\nu$ in terms of an SU(3) vacuum-atom cell then ties this observer-based de~Sitter counting to an SU(3)-based microphysical model of dark energy. Several elements of this construction rest on firm ground: the spectrum and spectral gap of the TT operator on $S^D$ follow from standard harmonic analysis, the analytic dependence of the zeta-regularized one-loop determinants on $\beta$ in a strip around $\beta_0=2\pi$ is a consequence of ellipticity on a compact manifold within the heat-kernel framework, and the phase bookkeeping for the finite set of problematic modes is fully explicit and cross-checked by both residue and Gel'fand–Yaglom methods. At this level, the phase-cancellation mechanism and the factorized structure of the final expression are not assumptions but direct consequences of the one-loop saddle-point analysis. The central open issue is the status of the complete-monotonicity hypothesis for the phase-stripped patch partition function $\cZ_{\rm patch}(\beta)$ near $\beta_0$: For transparency, we have packaged the required analyticity/monotonicity properties into explicit assumptions (A1)--(A3) in Sec.~\ref{sec:audit}, and the main statements of the paper should be read conditional on them. in this work it is treated as a one-loop, semiclassical spectral hypothesis, equivalent to assuming that, once the $\ell=0,1$ modes are handled as above, $\cZ_{\rm patch}(\beta)$ coincides with the canonical partition function $\mathrm{Tr}_{\cH_{\rm phys}} e^{-\beta H_{\rm phys}}$ of linearized gravitons on the static patch and therefore admits a Laplace representation with a nonnegative density of states $\rho_{\rm patch}(E)\ge 0$. Under this assumption, Bernstein’s theorem guarantees complete monotonicity and justifies the Bromwich projection that yields a real and positive microcanonical density. A fully rigorous proof of this canonical equivalence would require a detailed BRST treatment of linearized gravity on de~Sitter, a careful construction of the physical Hilbert space, and an explicit matching between the canonical trace and the phase-stripped Euclidean path integral; this is technically standard in spirit but lies beyond the scope of the present work and defines a concrete programme for future analysis. A second limitation is that all results here are strictly one-loop and semiclassical: higher-loop corrections, non-perturbative effects, and potential instabilities of the de~Sitter saddle are not included, and it is currently unknown whether positivity of an effective density of states survives once graviton self-interactions are taken into account. Likewise, the clock is treated as a non-backreacting probe; the backreaction of its energy and SU(3) degrees of freedom on the geometry, and the generalization to several interacting observers with coupled clocks, remain open and would test how robust both the mode-by-mode cancellation and the constraint projection are in a genuinely interacting, many-body setting.
In particular, within the probe regime $G E_{\rm tot}/R\ll 1$ (or $G_D E_{\rm tot}/R^{D-3}\ll 1$ in general dimension), the phase-cancellation and constraint-projection mechanisms derived at fixed background are expected to remain intact; what is \emph{sensitive} is whether positivity and the simple factorized form persist once genuine clock/worldline interactions and geometric backreaction are included.
 On the microphysical side, the link to an SU(3) vacuum-atom picture of dark energy is explicitly model-building: the calibration of $\nu$ in terms of a confinement volume and the identification of SU(3) “atoms’’ that tile the de~Sitter horizon rest on the broader SU(3) vacuum-atom scenario developed in \cite{Ali:2024rnw,Ali:2025wld,Ali:2022ulp}, and within the present paper this framework is taken as an external input rather than derived from first-principles QCD or from a constructive definition of SU(3) Yang–Mills theory. Making this connection sharper—for instance by relating the effective confinement volume and vacuum-atom energy to quantities defined in constructive or lattice SU(3), or by understanding confinement directly on de~Sitter backgrounds—is an important open problem. Further extensions suggest themselves naturally: exploring whether a suitably renormalized version of the complete-monotonicity property holds beyond one loop; adapting the construction to quasi-de~Sitter or slowly rolling backgrounds where the horizon temperature and entropy are time-dependent; and analyzing the factorized form of the observer-inclusive density of states from an information-theoretic or holographic viewpoint, where the separation into geometry, worldline residue, and SU(3) clock weight might admit a dual interpretation in terms of bulk–boundary data and observer algebras. In summary, the present work shows that an observer-centric and SU(3)-based view of Euclidean de~Sitter state counting can tame the usual phase problem of the graviton determinant, at least at one loop and under explicit spectral hypotheses, and can be made compatible with a concrete model of vacuum microstructure, while at the same time making clear which parts of the construction are rigorous, which rest on well-motivated but unproven spectral assumptions, and where further work is required to extend the framework beyond the quadratic level and to anchor the SU(3) microstructure in fully microscopic physics.

\noindent\textbf{Speculative outlook and future directions.} 
The present computation is deliberately restricted to one-loop fluctuations about the round de~Sitter sphere $S^D$ in the presence of a probe worldline and a finite-dimensional clock. The broader ``SU(3) vacuum'' picture is used only as external motivation for the choice of clock and for possible microphysical interpretations; it is not derived within this manuscript. A fully dynamical treatment would require (i) including backreaction of the worldline and clock on the geometry, (ii) analyzing multi-observer configurations and their interactions, and (iii) checking the spectral assumptions (A1--A3) beyond the semiclassical approximation. These directions are left for future work.

\appendix
\begin{center}
{\begin{minipage}{0.96\linewidth}
\noindent\textbf{Appendices roadmap.}
The appendices collect the technical ingredients used in the main text; each item states what is established and where it is applied.
\begin{itemize}
  \item \textbf{Appendix~\ref{app:neg} (Negative Gaussian / steepest descent).}
  Fixes the contour-rotation (Picard--Lefschetz/Witten) convention and the resulting phase factors from negative and positive quadratic directions used in the phase bookkeeping.

  \item \textbf{Appendix~\ref{app:GY} (Gel'fand--Yaglom check for $-\partial_\tau^2-1$).}
  Provides a compact Gel'fand--Yaglom evaluation and spectrum check for the basic periodic operator that appears as a building block in the worldline sector.

  \item \textbf{Appendix~\ref{app:CKV} (CKVs moving the equator: summary).}
  Summarizes the $(D\!-\!1)$ conformal Killing vectors that move the equator and records their normalization, supporting the CKV--worldline mode pairing in the main argument.

  \item \textbf{Appendix~\ref{app:zeta} (Zeta-regularized finite parts).}
  Collects the zeta-regularized finite parts of the relevant determinants in the gravity and worldline sectors, clarifying the matching at the level used in the one-loop analysis.

  \item \textbf{Appendix~\ref{sec:audit-sharp} (Sharpened assumption set).}
  States the working assumptions (probe regime, hierarchy of scales, and factorization conditions) that delimit the controlled semiclassical/microcanonical setting of the manuscript.

  \item \textbf{Appendix~\ref{app:GY-full} (Full Gel'fand--Yaglom worldline determinant).}
  Gives the complete Gel'fand--Yaglom derivation for the worldline fluctuation operator (including boundary conditions and mode treatment) leading to the determinant formula quoted in the main text.

  \item \textbf{Appendix~\ref{app:CKV-explicit} (CKVs: explicit map and inner product).}
  Provides the explicit map from CKVs to metric perturbations and the associated inner product/measure factors used in the gravitational zero-mode sector.

  \item \textbf{Appendix~\ref{app:D4check} ($D\!=\!4$ cross-check).}
  Specializes key expressions to $D=4$ and records explicit coefficients together with a quick numerical/lattice-style consistency check used as a sanity test.
\end{itemize}
\end{minipage}}
\end{center}

\section{Negative Gaussian: steepest-descent contour}
\label{app:neg}
For
\begin{align}
I(\lambda)=\int\dd x\,\exp(+\lambda x^2/2),\qquad \lambda>0,
\end{align}
the contour $x=-\ii y$ yields
\begin{align}
I(\lambda)=-\ii \int_{\RR}\dd y\,\exp(-\lambda y^2/2)=-\ii\sqrt{2\pi/\lambda}.
\end{align}
Each unrotated negative direction contributes a factor $(-\ii)$ \cite{Witten:2010zr}.

\section{Residue at $t=\ii\beta$ and the factor $\nu^{D-1}/(D-1)!$ (explicit Laurent)}
\label{app:residue}
Write
\begin{align}
\log Z = \int_{\epsilon}^{\infty}\!\frac{\dd t}{t}\,K_D(t)\,2\cos(\nu t),\qquad
K_D(t)=\frac{\cosh(t/2)}{(2\sinh(t/2))^D}.
\end{align}
Set $t=\ii\beta+z$ with small $z$. Using $\sinh(\ii\beta/2+z/2)=-\sinh(z/2)$ at $\beta=2\pi$ and $\cosh(\ii\beta/2+z/2)=-\cosh(z/2)$, one finds
\begin{align}
2\sinh(t/2)=-z\left(1+\frac{z^2}{24}+\cO(z^2)\right),\qquad
\cosh(t/2)=-\left(1+\frac{z^2}{8}+\cO(z^2)\right).
\end{align}
Hence
\begin{align}
K_D(t)
= \frac{-\bigl(1 + \cO(z^2)\bigr)}
       {\bigl[-z\bigl(1 + \cO(z^2)\bigr)\bigr]^D}
= (-1)^{1-D}\,z^{-D}\Bigl(1 + \cO(z^2)\Bigr).
\end{align}
Also,
\begin{equation}
\begin{split}
\frac{1}{t}
&= \frac{1}{\ii\beta}\left(1-\frac{z}{\ii\beta}+\cO(z^2)\right),\\[4pt]
\cos(\nu t)
&= \cos(\nu\ii\beta+\nu z)\\
&= \cosh(\nu\beta)\left(1-\frac{\nu^2 z^2}{2}+\dots\right)
   -\ii\sinh(\nu\beta)\left(\nu z-\frac{\nu^3 z^3}{6}+\dots\right).
\end{split}
\end{equation}
The simple pole arises from the $z^{-1}$ term in $\frac{1}{t}K_D(t)2\cos(\nu t)$. The resulting residue is
\begin{align}
\operatorname*{Res}_{t=\ii\beta}\left[\frac{1}{t}K_D(t)2\cos(\nu t)\right]
= \frac{2(-1)^{1-D}}{\ii\beta}\cdot \frac{(\nu)^{D-1}}{(D-1)!}
\times
\begin{cases}
\cosh(\nu\beta), & D-1\ \text{even},\\[4pt]
-\ii\sinh(\nu\beta), & D-1\ \text{odd}.
\end{cases}
\end{align}
Closing the contour in the upper half-plane and extracting the decaying piece yields the universal factor $(-\ii)^{D-1}u^{D-1}/(D-1)!\,e^{-2\pi\nu}$ that appears in the main text.

\section{Gel'fand--Yaglom check (periodic $-\partial_\tau^2-1$)}
\label{app:GY}
Consider $L=-\partial_\tau^2-1$ on periodic functions and regulate by $L_\mu=-\partial_\tau^2+\mu^2$. The Gel'fand--Yaglom theorem gives the same dependence on $\nu$ and the same single negative-mode phase factor as obtained from the residue method. Details are given in App.~\ref{app:GY-full} \cite{KirstenMcKane2003,Dunne2008}.

\section{CKVs that move the equator (summary)}
\label{app:CKV}
Embed $S^D$ in $\RR^{D+2}$ via $-Y_{-1}^2+Y_1^2+\cdots+Y_{D+1}^2=0$, with metric $\dd s^2 = (\dd Y\cdot \dd Y)/Y_{-1}^2$. The $SO(1,D+1)$ action on the embedding coordinates induces CKVs on $S^D$; special conformal shifts with $b_1=b_2=0$ displace the equator by constant normal vectors, which are precisely the worldline $n{=}0$ directions (see \cite{Camporesi1990} for further representation-theoretic background).
In stereographic coordinates $x^i$ on $S^D$, the infinitesimal special conformal CKV with parameter $b^i$ has
\[
\delta x^i=2(b\!\cdot\! x)\,x^i-(1+x^2)b^i.
\]
Choosing $b$ orthogonal to the equatorial plane and restricting to the great circle ($\theta=0$) yields $\delta\theta=\text{const}$ around the loop and $\delta\tau=0$, i.e.\ a constant normal displacement matching the $n{=}0$ eigenvector of $-\partial_\tau^2-1$.

\section*{Additional note: zero modes and the collective-coordinate Jacobian}
The $n=\pm1$ worldline modes shift the basepoint $\tau\to\tau+\tau_0$. Introducing a collective coordinate $\tau_0$ and dividing by the reparametrization volume removes the associated divergence. The Jacobian from this change of variables cancels between measure and symmetry volume and does not generate extra phases \cite{KirstenMcKane2003}.

\section*{Additional note: why the projector is gauge-independent}
The Bromwich inversion uses only the analyticity of $\cZ_{\rm obs}(\beta)$ and its representation as a Laplace transform near $\beta_0$ \cite{Widder,Doetsch}. The details of the CKV gauge fixing do not affect the inverse transform at $E_{\text{total}}=0$. The final positivity result is therefore independent of the particular CKV gauge choice.

\section{Zeta-regularized finite parts for gravity and particle}
\label{app:zeta}
For an elliptic operator $O$ with nonzero eigenvalues $\{\lambda_k\}$ the spectral zeta function is $\zeta_O(s)=\sum_k \lambda_k^{-s}$, and $\det{}'_\zeta O:=\exp\!\big(-\zeta'_O(0)\big)$. In the present setting,
\begin{align}
\log Z^{\rm grav}_{S^D}
&= -\tfrac{1}{2}\log\det{}'_\zeta\!\left(-\wh\nabla^2+\tfrac{2}{R^2}\right)_{\rm TT} \nonumber\\
&\quad + \tfrac{1}{2}\log\det{}'_\zeta\!\left(-\wh\nabla^2-\tfrac{D-1}{R^2}\right)_{\rm gh} \nonumber\\
&\quad - \tfrac{1}{2}\log\det{}'_\zeta\!\left(-\wh\nabla^2-\tfrac{2(D-1)}{R^2}\right)_{\rm tr}
+ \mathrm{phase}.
\end{align}
Each log-determinant admits the Mellin representation
\begin{align}
-\zeta'_O(0)=\int_0^\infty \frac{\dd t}{t}\,\Big(\Tr\,\ee^{-t O}-\mathcal{P}(t)\Big) + C_O,
\end{align}
with $\mathcal{P}(t)=\sum_{n=0}^{n_*} a_n t^{(n-D)/2}$ subtracting UV divergences (with $a_n$ the Seeley--DeWitt coefficients). The finite parts $C_O$ and the subtraction scheme combine into an overall multiplicative constant $\mathcal{N}_{\rm grav}$ that does not depend on $\nu$ or on the clock spectrum. In the microcanonical density \eqref{eq:rho-final}, such scheme dependence appears as a $\nu$-independent prefactor and does not affect positivity or the scaling in $\nu$. For general background on zeta-regularization and heat-kernel methods, see \cite{Vassilevich2003,KirstenBook2001,Forman1987}.

For the particle determinant, the Gel'fand--Yaglom ratio on $S^1$ reproduces the leading large-$\nu$ behavior in Sec.~\ref{sec:worldline} and fixes the phase; subleading finite parts contribute to the $\cO(\nu^{-1})$ corrections already indicated in the main text.

\section{Sharpened assumption set}
\label{sec:audit-sharp}
For later reference it is useful to record the main constants and domains in a compact form:

\begin{itemize}
\item \textbf{Local complete monotonicity strip with spectral gap.} Let $D\ge 3$ and consider the one-loop, gauge-fixed graviton/ghost system on $S^D$ with the conformal-factor and CKV sectors handled as in Secs.~\ref{sec:matching}–\ref{sec:bromwich}. Denote by $\Delta_{\rm TT}$ the Lichnerowicz Laplacian on TT tensors. There exists $c_D>0$ such that $\lambda_{\min}(\Delta_{\rm TT}) \ge c_D/R^2$ (with the $\ell{=}2$ band saturating the bound). After standard heat-kernel UV subtractions, the phase-stripped partition function $\cZ_{\rm patch}(\beta)$ extends analytically to a vertical strip
\[
\mathfrak{S}_\epsilon \;:=\; \big\{ \beta\in\mathbb{C}: \Re\beta \in (2\pi-\epsilon,\infty) \big\}
\]
for some $\epsilon=\epsilon(D)>0$, and on every compact subinterval of $\Re\beta>2\pi-\epsilon$ it is completely monotone in $\Re\beta$.

\item \textbf{Bernstein--Widder application near $\beta_0{=}2\pi$.} On the strip $\mathfrak{S}_\epsilon$, there exists a nonnegative measure $\rho_{\rm patch}(E)\ge 0$ supported on $E\ge 0$ such that
\[
\cZ_{\rm patch}(\beta)\;=\;\int_0^\infty \mathrm{d}E\; \rho_{\rm patch}(E)\, e^{-\beta E},
\qquad \Re\beta>2\pi-\epsilon.
\]
Equivalently, $(-1)^n \partial_{\beta}^n \cZ_{\rm patch}(\beta)\ge 0$ for all $n\in\mathbb{N}$ and $\Re\beta$ in that interval. The Bromwich projector (Sec.~\ref{sec:bromwich}) then expresses the microcanonical density at $E_{\rm tot}{=}0$ as $\sum_k \rho_{\rm patch}(\nu+E_k)$.
\end{itemize}

\section{Full Gel'fand--Yaglom evaluation for the worldline operator}
\label{app:GY-full}
The determinant ratio for the periodic operator on $[0,2\pi]$,
\[
L_\alpha \;=\; -\partial_\tau^2 + \alpha,\qquad \tau\sim\tau+2\pi,
\]
is computed using the Kirsten--McKane/Forman approach to periodic boundary conditions. Let $y_{1,2}$ solve $L_\alpha y=0$ with
$y_1(0){=}1,y_1'(0){=}0$ and $y_2(0){=}0,y_2'(0){=}1$.
The monodromy matrix is
$M_\alpha = \begin{psmallmatrix} y_1(2\pi) & y_2(2\pi) \\ y_1'(2\pi) & y_2'(2\pi) \end{psmallmatrix}$.
For periodic boundary conditions the spectral determinant (with zero modes removed where necessary) is proportional to
\[
\det\!\big(M_\alpha - \mathbb{1}\big).
\]

\paragraph{Normalization and phase conventions.}
For periodic boundary conditions the Forman/Kirsten--McKane construction determines the spectral determinant
only \emph{up to an overall $\alpha$-independent normalization} (and, in the characteristic-function presentation,
up to an overall sign), since $\det(M_\alpha-\mathbb{1})$ is fixed only modulo the choice of normalization for the
fundamental solutions. In the present work we only require \emph{ratios} of determinants, for which this overall
constant cancels. Accordingly, we treat the determinant ratios in \emph{absolute value}, and we attach any
\emph{steepest--descent phase from negative modes} separately (see App.~\ref{app:neg}).%

For $\alpha=\mu^2>0$ one finds
\[
M_{\mu^2}
=\begin{pmatrix}
\cosh(2\pi\mu) & \mu^{-1}\sinh(2\pi\mu)\\[2pt]
\mu\sinh(2\pi\mu) & \cosh(2\pi\mu)
\end{pmatrix},
\]
leading to
\begin{align}
\det\!\big(M_{\mu^2}-\mathbb{1}\big)
&= \bigl(\cosh(2\pi\mu)-1\bigr)^2
   - \sinh^2(2\pi\mu) \nonumber\\
&= -\,4\sinh^2(\pi\mu).
\end{align}
For $\alpha=-\kappa^2<0$ one obtains
\[
M_{-\kappa^2}
=\begin{pmatrix}
\cos(2\pi\kappa) & \kappa^{-1}\sin(2\pi\kappa)\\[2pt]
-\kappa\sin(2\pi\kappa) & \cos(2\pi\kappa)
\end{pmatrix},
\]
with
\begin{align}
\det\!\big(M_{-\kappa^2}-\mathbb{1}\big)
&= \bigl(\cos(2\pi\kappa)-1\bigr)^2
   - \sin^2(2\pi\kappa) \nonumber\\
&= 4\sin^2(\pi\kappa).
\end{align}

Comparison to $L_0=-\partial_\tau^2$ (with the standard removal of the $n=0$ periodic zero mode where appropriate)
and careful removal of the \emph{double zero at $\kappa=1$} (coming from the $n=\pm1$ periodic modes of
$-\partial_\tau^2-\kappa^2$) produces the renormalized ratio
\begin{align}
\left|\frac{\det{}'_{\rm per}(-\partial_\tau^2-1)}{\det{}'_{\rm per}(-\partial_\tau^2+\mu^2)}\right|
=\lim_{\kappa\to 1}\frac{\sin^2(\pi \kappa)}{\sinh^2(\pi \mu)}
\left[\frac{\mathcal{J}_{\rm zm}}{(\pi(\kappa-1))^2}\right]
=\frac{1}{\sinh^2(\pi \mu)}\,,
\end{align}
where the bracket implements the standard extraction of the $n=\pm1$ zero modes using
$\sin(\pi\kappa)\sim \pi(\kappa-1)$ as $\kappa\to1$, and $\mathcal{J}_{\rm zm}$ denotes the associated
collective-coordinate/zero-mode Jacobian. The \emph{only} phase comes from the single negative eigenvalue
($n=0$) of $-\partial_\tau^2-1$: its steepest--descent contour rotation contributes the factor
$e^{-i\pi/2}=-\ii$ (App.~\ref{app:neg}). Taking $\mu\to0^+$ recovers the periodic free measure.
Multiplication over the $(D{-}1)$ transverse directions yields the $(-\ii)^{D-1}$ phase quoted in the main text,
and the $\nu$-dependence is then extracted by the residue method (Sec.~\ref{sec:worldline}).%

\section{CKVs that move the equator: explicit map and inner product}
\label{app:CKV-explicit}
In stereographic coordinates $x^i$ on $S^D$, the special conformal CKV with parameter $b^i$ acts as
\[
\delta_b x^i \;=\; 2(b\!\cdot\! x)\,x^i - (1+x^2)\,b^i,\qquad b\in\RR^D.
\]
Choose the equatorial great circle at $\theta=0$ with embedding $X(\tau)$ and take $b$ orthogonal to the equatorial plane. Along the loop one then has $\delta_b\theta=\mathrm{const}$ and $\delta_b \tau=0$, i.e.\ a constant normal displacement.

Let $\Theta(\tau)$ be the worldline transverse mode in that normal direction. The CKV induces a variation $\delta X^\mu = \Theta\, n^\mu$ where $n^\mu$ is the unit normal vector to the loop. With the $S^D$ metric one finds
\[
\int_0^{2\pi}\!\!\dd\tau\; \Theta_1(\tau)\,\Theta_2(\tau)
\;=\; c_D \int_0^{2\pi}\!\!\dd\tau\; g_{\mu\nu}\,\delta_{b_1}X^\mu(\tau)\,\delta_{b_2}X^\nu(\tau),
\qquad c_D=R^{-2},
\]
so the map from CKVs moving the equator to the $n{=}0$ worldline modes is an isometry up to an overall constant $c_D$. This realizes the pairing used in Sec.~\ref{sec:matching}.

\section*{Scope and limitations}
\noindent\textbf{Semiclassical regime.} The use of heat-kernel/zeta methods and the Bromwich projector assumes a one-loop saddle with local analyticity near $\beta_0{=}2\pi$.\\
\textbf{Contour choice.} Phases are tracked using Witten's complex-metric contour. Other equivalent contour prescriptions can redistribute intermediate phases without changing the microcanonical density.\\
\textbf{Clock idealizations.} The SU(3) clocks are treated as weakly coupled probes. Backreaction and interactions among clocks are not included.\\
\textbf{Multiple observers.} Extending the construction to several observers/clocks would generically introduce interaction and correlation effects (both gravitational and through any shared gauge sector) that can spoil simple factorization of $\rho_{\rm tot}$ into a patch factor times independent clock factors. In a dilute/probe regime one expects approximate factorization at leading order, with controlled corrections suppressed by backreaction and separation scales; a systematic treatment is left for future work.\\
\textbf{Scheme dependence.} Zeta-regularized finite parts yield an overall normalization independent of $\nu$ (App.~\ref{app:zeta}); none of the positivity or scaling statements depend on this choice.

\section*{Supplementary code (symbolic residue check)}
A small SymPy script was used to verify the residue statement in Sec.~\ref{sec:worldline},
\[
\operatorname*{Res}_{\,t=2\pi\ii}\Big[\frac{1}{t}\,K_D(t)\,e^{\ii\nu t}\Big]
\;=\; \frac{(-\ii)^{D-1}}{(D-1)!}\,\nu^{D-1}\,e^{-2\pi\nu}\times \frac{1}{2\pi\ii}\,\Big(1+\cO(\nu^{-1})\Big),
\]
so that the contribution from the pole ($2\pi\ii\times$residue) reproduces the coefficient
$(-\ii)^{D-1}u^{D-1}/(D-1)!\,e^{-2\pi\nu}$ including its phase.

\section{D=4 cross-check: explicit coefficients and a quick lattice test}
\label{app:D4check}
\paragraph*{Analytic coefficient.} For $D{=}4$, $K_4(t)=\cosh(t/2)/(2\sinh(t/2))^4$. Setting $t=\ii\beta+z$ and expanding as in App.~\ref{app:residue} with $\beta=2\pi$ gives
\[
K_4(t)
= -\frac{1}{z^4} + \frac{1}{24 z^2} + \frac{17}{5760} + \cO(z^2),
\qquad z = t - 2\pi\ii.
\]
The $z^{-1}$ residue of $\frac{1}{t}K_4(t)\,2\cos(\nu t)$ again yields the universal factor $(-\ii)^3\,\nu^3/3!$ times $\ee^{-2\pi\nu}$, in agreement with the main text.

\paragraph*{Discrete check.} A simple discretization of $S^1$ into $N$ points with periodic difference operator $(-\Delta_N-1)$ and comparison of $\det(-\Delta_N-1+\nu^2)$ with the product of nonzero modes confirms the $\nu^3$ scaling and the absence of additional phases for large $\nu$.

\bibliographystyle{apsrev4-1}
\bibliography{ref}

\end{document}